\pdfoutput=1
\documentclass[aps,prb,twocolumn,showpacs,preprintnumbers,amsmath,amsfonts,amssymb,groupedaddress,longbibliography,floatfix,reprint]{revtex4-1}

\usepackage[english]{babel}
\usepackage[utf8]{inputenc}

 \usepackage{geometry}
\usepackage{todonotes}
\usepackage[caption=false]{subfig}
\usepackage{graphicx}

\usepackage{amsfonts}
\usepackage{amsmath}
\usepackage{amssymb}
\usepackage{mathtools}

\usepackage[]{xcolor}
 \definecolor{lightblue}{RGB}{0,0,255}
 \definecolor{darkblue}{RGB}{0,0,128}
 \definecolor{orange}{RGB}{255,127,0}
 \definecolor{green}{RGB}{128,128,0}
 
\usepackage{siunitx}
\usepackage[
bookmarks=true,
bookmarksopen=false,
bookmarksnumbered=true,
breaklinks=true,
colorlinks=true,
linkcolor=blue,
anchorcolor=black,
citecolor=black,
filecolor=black,
menucolor=black,
urlcolor=black,
pdfpagemode=UseOutlines,
pdftitle=Exotic Superconductivity Through Bosons in a Dynamical Cluster Approximation,
pdfsubject=Exotic Superconductivity Through Bosons in a Dynamical Cluster Approximation,
pdfauthor=Thomas Bilitewski
]{hyperref}
 
\newcommand{\vect}[1]{\mathbf{#1}}

\DeclareSymbolFont{bbold}{U}{bbold}{m}{n}
\DeclareSymbolFontAlphabet{\mathbbold}{bbold}

\makeatletter
\newlength{\sfp@hseplen}\newlength{\sfp@vseplen}
\define@cmdkey{subfigpos}[sfp@]{pos}[ul]{}
\define@cmdkey{subfigpos}[sfp@]{font}[\small]{}
\define@cmdkey{subfigpos}[sfp@]{vsep}[2\baselineskip]{\setlength{\sfp@vseplen}{\sfp@vsep}}
\define@cmdkey{subfigpos}[sfp@]{hsep}[10pt]{\setlength{\sfp@hseplen}{\sfp@hsep}}
\newcommand{\subfigimg}[3][,]{%
  \setkeys{Gin,subfigpos}{pos,font,vsep,hsep,#1}
  \setbox1=\hbox{\includegraphics{#3}}
  \ifnum\pdfstrcmp{\sfp@pos}{ul}=0
    \leavevmode\rlap{\usebox1}
    \rlap{\hspace*{\sfp@hsep}\raisebox{\dimexpr\ht1-\sfp@vsep}{\sfp@font{#2}}}
    \phantom{\usebox1}
  \else\ifnum\pdfstrcmp{\sfp@pos}{ur}=0
    \leavevmode\usebox1
    \llap{\raisebox{\dimexpr\ht1-\sfp@vsep}{\sfp@font{#2}}\hspace*{\sfp@hsep}}
  \else\ifnum\pdfstrcmp{\sfp@pos}{lr}=0
    \leavevmode\usebox1
    \llap{\raisebox{\sfp@vsep}{\sfp@font{#2}}\hspace*{\sfp@hsep}}
  \else
    \leavevmode\rlap{\usebox1}
    \rlap{\hspace*{\sfp@hseplen}\raisebox{\sfp@vsep}{\sfp@font{#2}}}
    \phantom{\usebox1}
  \fi\fi\fi
}
\makeatother

\begin{document}

\title{Exotic Superconductivity Through Bosons in a Dynamical Cluster Approximation }
\author{Thomas \surname{Bilitewski}}
\email{tb494@cam.ac.uk}
\affiliation{T.C.M. Group, Cavendish Laboratory, J.J. Thomson Avenue, Cambridge CB3 0HE, United Kingdom}
\author{Lode \surname{Pollet}}
\affiliation{Department of Physics, Arnold Sommerfeld Center for Theoretical Physics, and Center for Nanoscience, Ludwig-Maximilians-Universit{\"a}t M{\"u}nchen, Theresienstra{\ss}e 37, 80333 Munich, Germany}

\date{\today}
\begin{abstract}
We study the instabilities towards (exotic) superconductivity of mixtures of spin-$1/2$ fermions coupled to scalar bosons on a two-dimensional square lattice with the Dynamical-Cluster-Approximation (DCA) using a numerically exact continuous-time Monte-Carlo solver. The Bogoliubov bosons provide an effective phononic bath for the fermions and induce a non-local retarded interaction between the fermions, which can lead to (exotic) superconductivity. Because of the sign problem the biggest clusters we can study are limited to $2 \times 2$ in size, but this nevertheless allows us to study the pairing instablilities, and their possible divergence, in the  $s$-
and $d$ -wave channels as well as the competition with antiferromagnetic fluctuations. At fermionic half-filling  we find that $d$-wave is stable when the mediated interaction by the bosons is of the same order as the bare fermionic repulsion. Its critical temperature can be made as high as the maximum one for $s$-wave, which opens perspectives for its detection in a cold atom experiment.
\end{abstract}


\maketitle

\makeatletter
 \newpage

\section{\label{sec:intro}Introduction}
Mixtures of bosons and fermions are ubiquitous: They are fundamental in particle physics, where bosons are the carriers of forces between fermionic matter particles. In condensed matter systems, they appear in the context of superconductivity, where the conventional pairing mechanism consists of phonons inducing an effective retarded attractive interaction between electrons leading to the formation of Cooper-pairs, as well as in mixtures of $^3$He and $^4$He in which the inter- and intra-isotope interactions are of comparable magnitude. Cold-atom systems are uniquely suited to simulate this physics as they allow fine experimental control over the interactions as well as the tunelling amplitudes \cite{Bloch2008,Ketterle_2008}.

The optical lattice system does not have phonons. Nevertheless, pairing mechanisms with cold fermions in an optical lattice can be investigated by quantum simulation. Bosons deep in the superfluid phase have a linear dispersion~ \cite{Wang2005} which can play the role of phonons as in conventional superconductors (when fermions form an electron gas) but, as we will see, they can equally well couple to spin density wave fluctuations (relevant when the charge of the fermions is near localization). Spin density wave fluctuations induce an attractive interaction between electrons in a spin-singlet, with the pair wavefunction changing sign between different regions of the Fermi surface~\cite{Scalapino_1986,Abanov_2001,Abanov_2003,Chubukov_2013,Metlitski_2010a,Metlitski_2010b,Berg_2011, Sachdev_2009,Sachdev_2012, Scalapino_2012}. It plays a prominent role in the cuprates (where it correctly predicts $d$-wave pairing), the iron superconductors and heavy fermion materials~\cite{Zhai_2009,Moon_2010,Kuroki_2008,Mazin_2008,Scalapino_2009,Chubukov_2010,Paglione_2010,Johnston_2010,Gegenwart_2008,Nair_2010}. The fermionic Hubbard repulsion and the lattice dispersion can, in principle, be changed in a cold-atom experiment and allow to systematically investigate the interplay of both pairing mechanisms.

Quantum degenerate mixtures of bosonic and spin-polarized fermionic species have first been realized experimentally with $^{23}\mathrm{Na}$ and $^{6}\mathrm{Li}$ \cite{Hadzibabic_2002}. Since then a variety of different species combinations have been employed in experiments and these systems have been studied extensively
\cite{Buchler_2003,Lewenstein_2004,Buchler_2004,Cramer_2004,Roth_2004,Silber_2005,Ospelkaus_2006a,Ospelkaus_2006b,Guenter2006,Powell_2005,Ni_2008,Best2009,Pollet_2006,Pollet_2008,Parish_2008, Fratini_2010,Hansen_2011,Hara_2011,Yu_2012,Pieri_2013,Pierbiagio_2015}. 
Recently, using $^{6}\mathrm{Li}$ and $^{7}\mathrm{Li}$ a system has been realised experimentally for the first time in which both bosonic and spinful fermionic species are superfluid \cite{Ferrier_2014}, giving rise to induced interactions in the bosonic sector due to excitations of the fermionic superfluid \cite{Zheng_2014,Kinnunen_2015}. Another promising candidate for experiments with spinful mixtures are $^{23}\mathrm{Na}$ and $^{40}\mathrm{K}$ in which a large number of interspecies Feshbach resonances at experimentally accessible magnetic fields have been identified enabling the tuning of the interspecies interactions and the simulation of boson-induced interactions between the fermions \cite{Park_2011,Park2012} and the creation of stable fermionic Feshbach molecules \cite{Park_2015}.

Theoretically, the interactions mediated by superfluid bosons ~\cite{Bijlsma2000} were studied in the regime where the sound velocity of the bosons is fast compared to the Fermi velocity and found to be attractive and capable of overcoming a sufficiently weak repulsive fermi interactions leading to $s$-wave pairing both in the continuum \cite{Bijlsma2000,Heiselberg2000,Modugno_2002,Viverit2002,Viverit_2002a,Matera_2003,Wang_2006,Kalas_2008,Zwerger_2009} and in optical lattices \cite{Kagan_2004,Illuminati_2004,Wang2005}. Recently, the phase diagram of Bose-Fermi mixtures in 3 dimensional optical lattices has been studied numerically within the DMFT-formalism \cite{Anders2012} and analytically in a mean-field treatment \cite{Bukov2014} finding such phases as charge-density waves, superfluidity in either or both sectors, and supersolids. In 2 dimensional lattices the competition between $s$- and $d$-wave superfluidity and antiferromagnetic phases was investigated using the functional renormalization group \cite{Mathey2006}.

In this work we restore some momentum fluctuations compared to the aforementioned DMFT study by studying mixtures in the Fermi liquid regime in a DCA framework in two dimensions and monitor the pairing susceptibilities. We assume that the bosons are deep in the superfluid phase which allows to find (exotic) pairing channels (unlike single-site DMFT) and the interplay with anti-ferromagnetic fluctuations but it rules out charge density wave order. We find that the bosonic condensate can enhance both $s$- and $d$-wave pairing. Just as in conventional superconductors $s$-wave pairing is possible for weakly repulsive bare fermions. However, as the lattice effects and the bare Hubbard repulsion grow in importance, $d$-wave takes over and can be the dominant pairing channel.  Atomic Bose-Fermi mixtures thus effectively display two different pairing mechanisms, relevant for superconductors. \\

\section{\label{sec:model}Model}
Our system is described by the fermionic action $ S  = S_{\rm f} + S_{\rm ret}$, with
\begin{align}\label{eq:action}
S_{\rm f} &= \int_0^{\beta} {d\tau \, \sum_{\left< i,j \right>,\sigma}{ \bar{c}_{i}^{\sigma}(\tau) \left[\delta_{i,j} \left(\partial_{\tau}-\mu_{\rm f}+n_0 U_{\rm bf}\right) - t_{\rm f}  \right] c_{j}^{\sigma}(\tau)}  } \notag\\
    & \qquad \qquad  + U_{\rm ff} \sum_i{n^{f}_{i,\uparrow}(\tau) \, n^{f}_{i,\downarrow}(\tau) } \\
S_{\rm ret} &= - \frac{n_0 \, U^{2}_{\rm bf}}{2} \iint_0^{\beta}{d\tau_1 d\tau_2 \,\sum_{i,j}}n_{i}^{f}(\tau_1)\, D_{ij}(\tau_1-\tau_2)\, n_{j}^{f}(\tau_2) \notag .
\end{align}
It describes a fermionic Hubbard model with hopping amplitude $t_{f}$, on-site repulsion $U_{\rm ff}$ and chemical potential $\mu_{f}$ coupled to bosons deep in the condensed phase with condensate density $n_0$ via an on-site density-density coupling $U_{\rm bf}$. The bosons are treated in the Bogoliubov approximation and subsequently integrated out, giving rise to the chemical potential shift $n_0 U_{\rm bf}$ and  the non-local retarded density-density action term $S_{\rm ret}$ with kernel 
\begin{equation}\label{eq:ret_int_kernel}
 D(i-j,\tau)=  \iint_{B.Z.}{\frac{d^{2}k}{(2\pi)^{d}} \,  e^{i\vect{k} (\vect{r}_i - \vect{r}_j)} \,   \frac{e^{E_{\vect{k}} \tau}+e^{E_{\vect{k}} (\beta-\tau)} }{e^{\beta E_{\vect{k}}} -1 } \, \frac{|\bar{\varepsilon}_{\vect{k}}|}{E_{\vect{k}}} }
\end{equation}
where $ E_{\vect{k}} = \left[ \bar{\varepsilon}_{\vect{k}}^{2}+ 2 \bar{\varepsilon}_{\vect{k}} n_0 U_{\rm bb} \right]^{1/2} $ is the dispersion of the Bogoliubov quasi-particles, $\bar{\varepsilon}_k = \varepsilon_{\vect{k}} - \varepsilon_{\vect{0}}$ the lattice dispersion of the bare bosons, shifted to be positive, and $U_{\rm bb}$ the repulsion between the bare bosons on the lattice.  The chemical potential $\mu_b$ was fixed to give unit filling for the bosons (its density is unimportant in the superfluid regime). The explicit construction from the underlying Hamiltonian is performed in Appendix~\ref{app:eff_action_construction}. This treatment neglects any back-action of the fermions on the bosons, in particular, bosonic charge order is excluded, which might be mediated from the fermionic sector by the density-density coupling between bosons and fermions. However, for fast bosons (amounting to $t_{\rm b} > t_{\rm f}$) which we consider, charge-density ordering was found to be suppressed and superconductivity to be favored in DMFT \cite{Anders2012}.

In the instantaneous limit, {\it i.e.}, the case of a very high bosonic speed of sound compared to the Fermi velocity as it can be made in the $\mathrm{NaK}$ system~\cite{Wang2005}, this simplifies to
$D_{\rm inst}(\vect{k}) = \frac{1}{ n_0 U_{\rm bb} }  c(\vect{k}) = \frac{1}{ n_0 U_{\rm bb} }  \left\{1 + {\xi}^{2} \left[4 -  \cos(k_x) - \cos(k_y)  \right]  \right\}^{-1}$, 
with $\xi = \sqrt{t_{\rm b}/(2 U_{\rm bb})}$ as obtained in \cite{Mathey2006}. The induced on-site interaction is given by $W=\frac{U_{\rm bf}^{2}}{ U_{\rm bb} } \sum_{\vect{k}} c(\vect{k}) $, thus scaling as 
 $V=U_{\rm bf}^{2}/ U_{\rm bb}$. The bosonic dispersion $E_k$ can only have an influence on the fermionic system if $\xi$ is comparable to the lattice spacing, which is satisfied for $\mathrm{NaK}$. \\

%
\section{\label{sec:method}Methodology}
To numerically study the action Eq.~(\ref{eq:action}) we use the dynamical cluster approximation (DCA) \cite{Hettler1998,Jarrell_2001a,Maier2005}, which maps the many-body problem onto a self-consistently embedded cluster impurity problem. Its action $S = S_{\rm 0} + S_{\rm int}$ reads 
\begin{align}\label{eq:cluster_action}
S_{\rm 0} &= \int_0^{\beta} {d\tau \, \sum_{i,j,\sigma}{ \bar{c}_{i}^{\sigma}(\tau_1) \mathcal{G}_{\sigma,i,j}^{-1}(\tau_1-\tau_2) c_{j}^{\sigma}(\tau_2)}  } \\
S_{\rm int } &= U_{\rm ff} \sum_i{n^{\rm f}_{i,\uparrow}(\tau) \, n^{\rm f}_{i,\downarrow}(\tau) } + S_{\rm ret}
\end{align}
Here, the sums run over the cluster degrees of freedom, $\mathcal{G}$ is the unknown cluster-excluded Green's function of the impurity problem which has to be determined self-consistently. The local Hubbard interaction $U_{\rm ff}$ remains unaffacted by the DCA mapping, but the non-local induced interaction is coarse-grained, {\it i.e.}, $D_{ij}$ in $S_{\rm ret}$ is replaced by $\bar{D}_{ij}$, which is the Cluster-Fourier transform of the coarse-grained interaction kernel $D(i-j)$.

To solve the impurity problem, we use a generalized weak-coupling solver \cite{Rubtsov2004,Rubtsov2005} to include the non-local phononic degrees of freedom similar as in Ref.~\cite{Assaad2007}. We are limited to $2 \times 2$ clusters because of the sign problem, which also occurs at half filling because the induced interactions, $\bar{D}_{i,j}(\tau)$ (Eq.~\ref{eq:ret_int_kernel}), do not have a definite sign. Results of the average sign for characteristic values of the parameters and temperatures used in our simulations are shown in Appendix~\ref{app:sign}. Unfortunately, $2 \times 2$ clusters are known to overestimate the $d$-wave transition temperature and to strongly suppress antiferromagnetic phases \cite{Maier2005,Maier2005a}. Bigger clusters are highly desired to take the Kosterlitz-Thouless transitions properly into account, but are out of reach.

To obtain transitions to a superconducting state we need to consider the two-particle Green's function for opposite spin pairing in the particle-particle channel
\begin{eqnarray}
 \chi(q,k,k^{\prime}) &= &\int_{0}^{\beta}\int_{0}^{\beta}\int_{0}^{\beta}\int_{0}^{\beta}{d\tau_1 d\tau_2 d\tau_3 d\tau_4} \\
   {}                      &\times & e^{i \left[(\omega_n +\nu)\tau_1 - \omega_n \tau_2 + \omega_{n^{\prime}} \tau_3 - (\omega_{n^{\prime}}+\nu)\tau_4 \right] } \left< T_{\tau} P(\cdot) \right> \nonumber \\
   P( \cdot )  & = & c^{\dagger}_{\vect{k}+\vect{q},\sigma}(\tau_1)\, c^{\dagger}_{-\vect{k},-\sigma}(\tau_2) \, c_{-\vect{k^{\prime}},-\sigma}(\tau_3) \, c_{\vect{k^{\prime}}+\vect{q},\sigma}(\tau_4)  \nonumber 
 \end{eqnarray}
where we adopt the notation $k=(\vect{k}, i\omega_n)$, $k^{\prime}=(\vect{k^{\prime}}, i\omega_{n^{\prime}})$ and $q= (\vect{q},i\nu_n)$ \cite{Abrikosov1975}.
For transitions to states of lower symmetry than the lattice, the Green's function needs to be projected according to
\begin{equation}
 \Pi_{g,g}(q,k,k^{\prime}) =g(\vect{k}) \chi(q,k,k^{\prime}) g(\vect{k^{\prime}})
\end{equation}
where $g(\vect{k})$ is the form-factor for the corresponding symmetry.
The pairing susceptibility is then given by
\begin{equation}\label{eq:pairing_susc}
 P_g(q,T) = \frac{T^{2}}{N_c^{2}} \sum_{K,K^{\prime}}{\overline{\Pi}_{g,g}(q,K, K^{\prime})} \, .
\end{equation}
where $\overline{\Pi}$ is the coarse-grained version of $\Pi_{g,g}$ obtained by inverting the coarse-grained Bethe-Salpeter equation as described in \cite{Jarrell_2001a}.
The calculation of the corresponding quantity for pairing in the particle-hole channel, relevant for transitions to an antiferromagnetic state, is similar and explained in detail in \cite{Maier2005}. \\ 

\section{\label{sec:results}Results}
\begin{figure} 
\centering 
\subfloat{\subfigimg[hsep=0.0\textwidth,vsep=0.0\textwidth,width=0.249\textwidth,pos=ul]{\bf (a)}{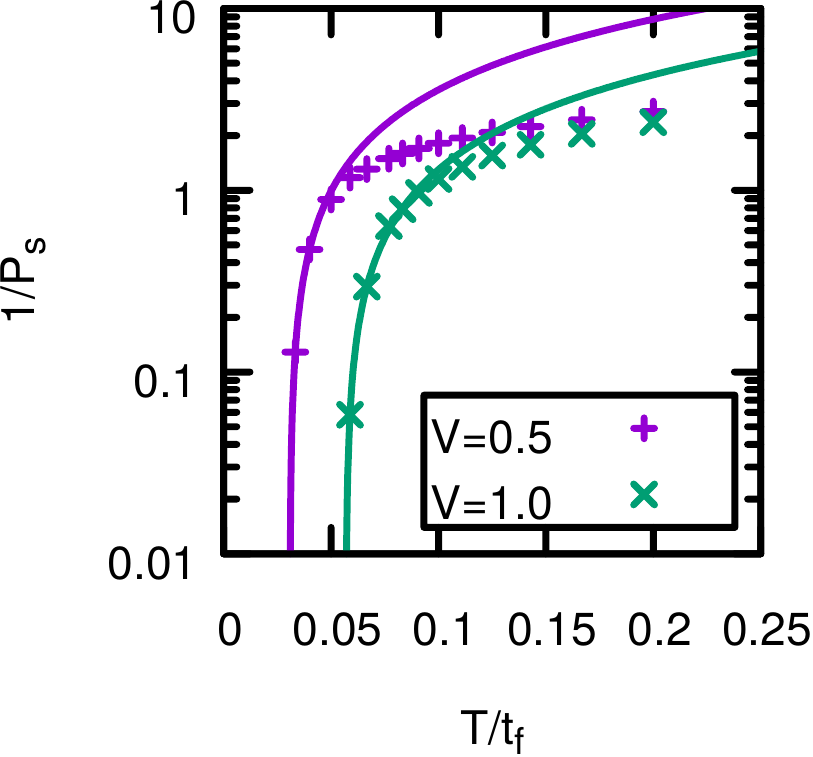}}
\subfloat{\subfigimg[hsep=0.0\textwidth,vsep=0.0\textwidth,width=0.249\textwidth,pos=ul]{\bf (b)}{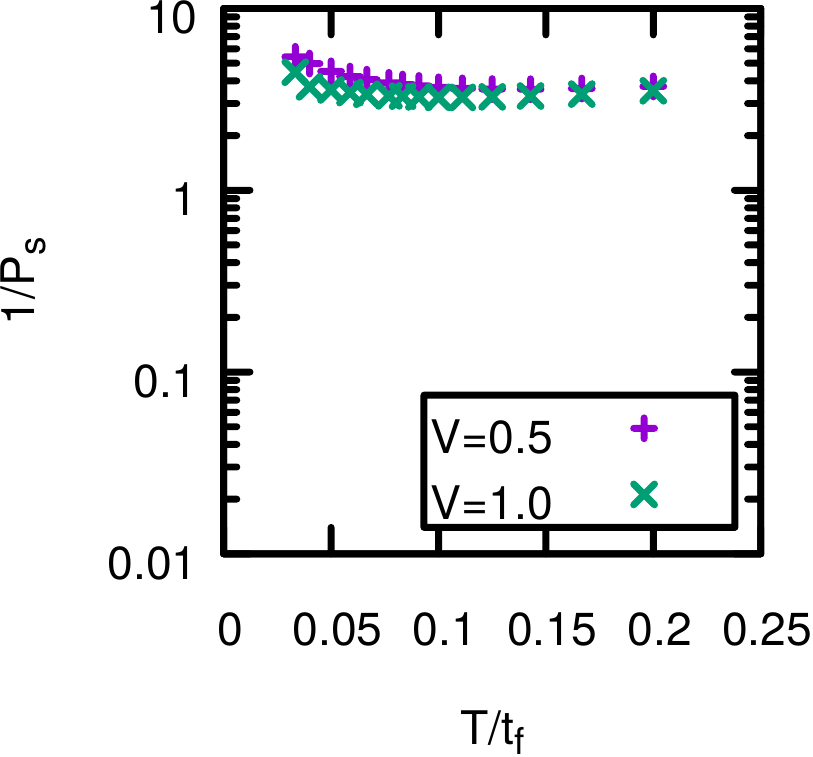}} 

\subfloat{\subfigimg[hsep=0.0\textwidth,vsep=0.0\textwidth,width=0.249\textwidth,pos=ul]{\bf (c)}{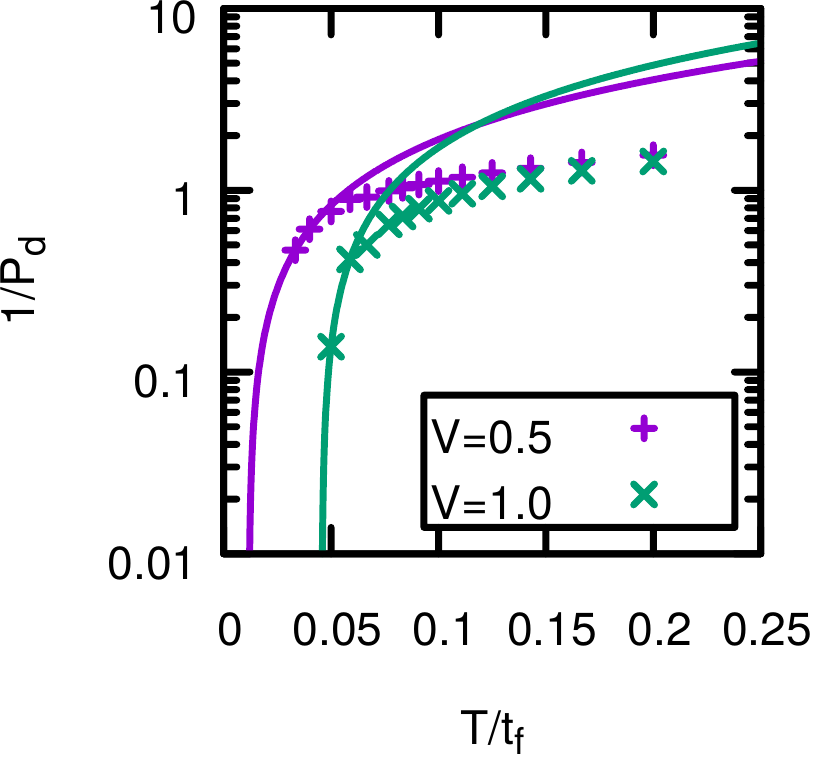}}
\subfloat{\subfigimg[hsep=0.0\textwidth,vsep=0.0\textwidth,width=0.249\textwidth,pos=ul]{\bf (d)}{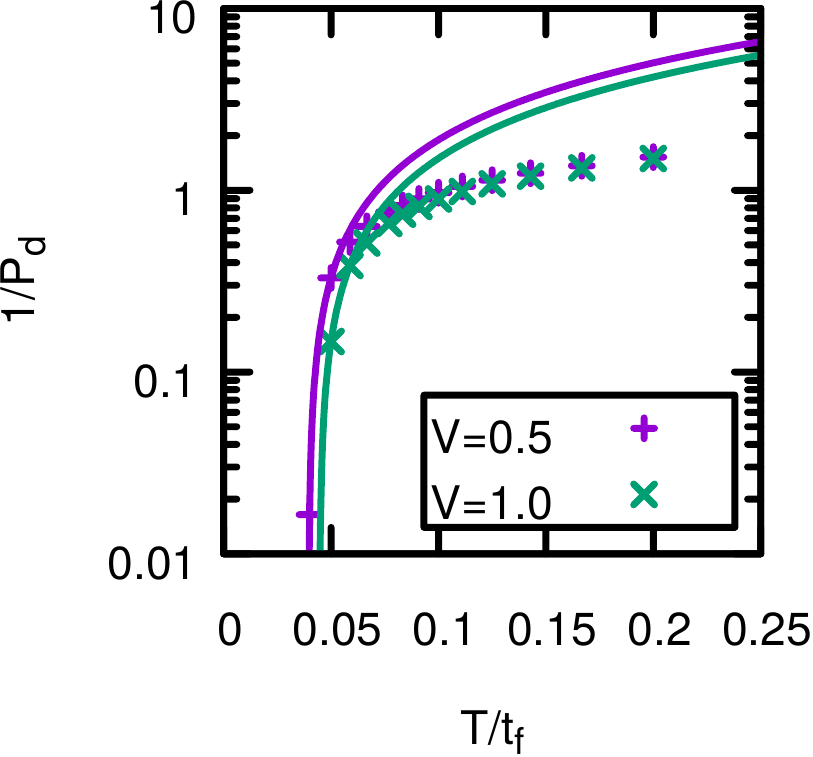}} 
  \caption{Inverse pairing field susceptibility $1/P_g$ on a logarithmic scale for $g = s(d)$-wave symmetry at the top (bottom) at fixed $U_{\rm ff}= 0 (1)$ at the left (right) and fermionic half filling as a function of temperature for different coupling strength $U_{\rm bf}$ to the bosons corresponding to $V=0.5,1.0$ for the bosonic parameters $U_{\rm bb}/t_{\rm f}=5$ and $t_{\rm b}/t_{\rm f}=10$. The vanishing of $1/P_g$ signals the transition to a superconducting state of the corresponding symmetry. The solid lines are linear fits to the lowest temperature points and are used to derive $T_c$.
\label{fig:transition_fast}}
\end{figure}
We now turn to the investigation of the fermionic instabilities, which is reflected in the behavior of susceptibilities in the particle-hole and particle-particle channels, specifically the antiferromagnetic susceptibility $\chi_{\rm AF}$ at momentum transfer $q=((\pi,\pi),0)$ and the pairing field susceptibility $P_g$ as defined in Eq.~(\ref{eq:pairing_susc}) for the dominant momentum transfer $q=(\vect{0},0)$ in the case of the $s$-, extended $s$-, $p$- and $d$-wave symmetry \cite{Jarrell_2001a,Maier2005a}. The divergence of the pairing field signals the phase transition to the state of the corresponding symmetry. To obtain the respective transition temperature we fit a linear function to the pairing fields in the low-temperature region as is appropriate for the mean-field nature of DCA close to $T_c$ \cite{Maier2005,Maier2005a}. At half filling, we only found transitions to $s$- and $d$-wave, as well as strong antiferromagnetic correlations, whose competition is the focus of what follows.

The $s (d)$-wave pairing fields with the linear fits are shown in the upper (lower) panel of Fig.~(\ref{fig:transition_fast}).
In the free model $(U_{\rm ff} = U_{\rm bf}=0)$ there is no pairing in either channel. When increasing $U_{\rm bf}$ (but keeping $U_{\rm ff}=0$), competing $s$- and $d$-wave instabilities develop as seen in the left panel. For $U_{\rm bf}/t_{\rm f}=1.58$, {\it i.e.}, $V/t_{\rm f}=0.5$, we observe that the extrapolated $s$-wave transition temperature is higher than the $d$-wave one and $s$-wave pairing is thus the dominant channel. For weak $V$ the mechanism for induced $s$-wave superconductivity is the same as in BCS theory. Upon increasing  $V$, the $s$-wave gap becomes stronger and we enter the regime where superconductors are routinely described by the Migdal-Eliashberg theory \cite{Mig_58,Eli_60,Carbotte_1990,Wang_2006,Springer_Superconductivity_ep,Bauer_2011,Margine_2013}. The momentum dependence of the bosonic dispersion $E_k$ is here unimportant but retardation effects do matter in general. A selfconsistent treatment of the bosons ({\it e.g,}, via a damping term) is left for future work.

Upon increasing $U_{\rm ff} $ $s$-wave-pairing is suppressed and $d$-wave is the dominant instability as seen in the right panel for $U_{\rm ff}/t_{\rm f}=1$ where no $s$-wave instability is found for the range of $V$ shown. This is easily understood in the instantaneous limit where $W$ scales with $V=U_{\rm bf}^{2}/U_{\rm bb}$ and for $|W|<|U_{\rm ff}|$ no $s$-wave pairing is possible.

\begin{figure}
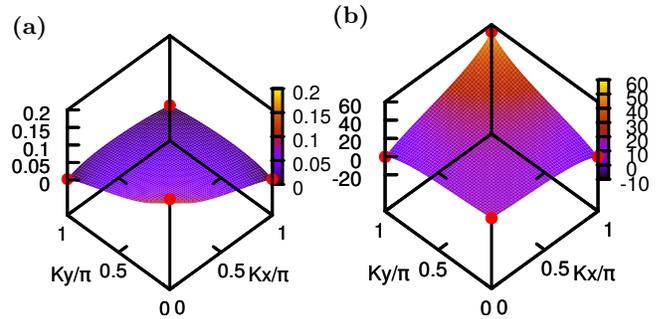

\centering 
\subfloat{\subfigimg[hsep=0.0\textwidth,vsep=0.0\textwidth,width=0.24\textwidth,pos=ul]{\bf (a)}{{./figures/plot_gamma_momentum_Uff_0.2_Ubf_3.16_beta_5_crop}.pdf}}
\hfill
\subfloat{\subfigimg[hsep=0.0\textwidth,vsep=0.0\textwidth,width=0.24\textwidth,pos=ul]{\bf (b)}{{./figures/plot_gamma_momentum_Uff_5_Ubf_2.24_beta_5_crop}.pdf}} 
\caption{Pairing vertex in the particle-particle channel $\Gamma^{PP}(\vect{K}-\vect{K^{\prime}})$ as a function of the relative momentum for $U_{\rm ff}/t_{\rm f}=0.2 $, $V/t_{\rm f}=2$ in (a) where $s$-wave pairing dominates showing a peak at $\vect{K}-\vect{K^{\prime}} =(0,0)$ and for  $U_{\rm ff}/t_{\rm f}=5 $, $V/t_{\rm f}=1$ in (b) where $d$-wave pairing dominates displaying a clear peak at $\vect{K}-\vect{K^{\prime}} =(\pi,\pi)$.
\label{fig:pairing_vertex}}
\end{figure}

The competition between $s$- and $d$-wave pairing is also clearly reflected in the momentum structure of the pairing vertex shown in Fig.~(\ref{fig:pairing_vertex}) in (a) for a situation where $s$- wave dominates and in (b) where $d$-wave is dominant. In the case of $s$-wave pairing the vertex peaks at a zero momentum transfer leading to a homogeneous real space structure whereas for $d$-wave a strong peak develops for a momentum transfer $\vect{K}-\vect{K^{\prime}} =(\pi,\pi)$ leading to oscillatory behaviour in real space.

\begin{figure} 
\centering 
\subfloat{\subfigimg[hsep=0.0\textwidth,vsep=0.0\textwidth,width=0.24\textwidth,pos=ul]{\bf (a)}{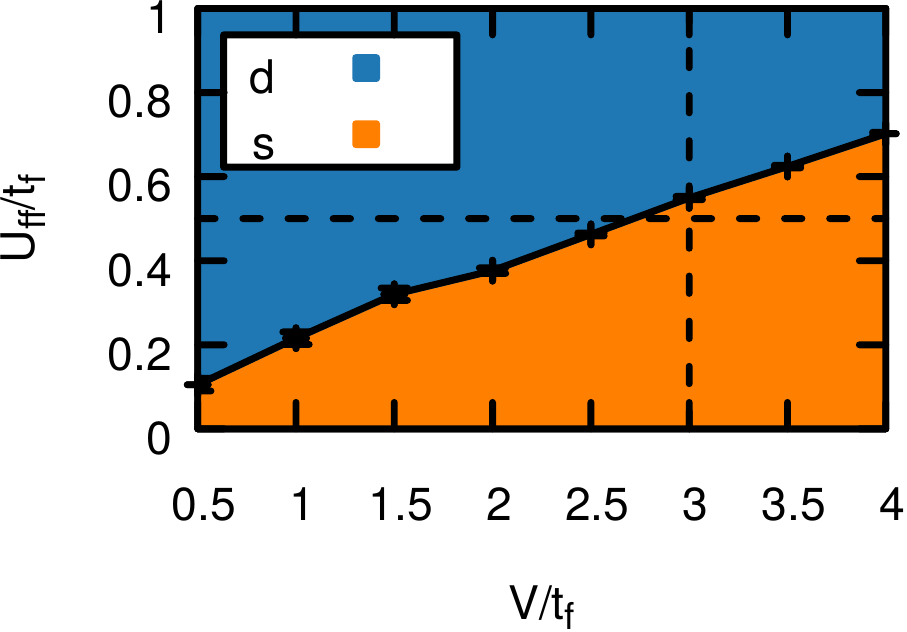}}
\hfill
\subfloat{\subfigimg[hsep=0.0\textwidth,vsep=0.0\textwidth,width=0.24\textwidth,pos=ul]{\bf (b)}{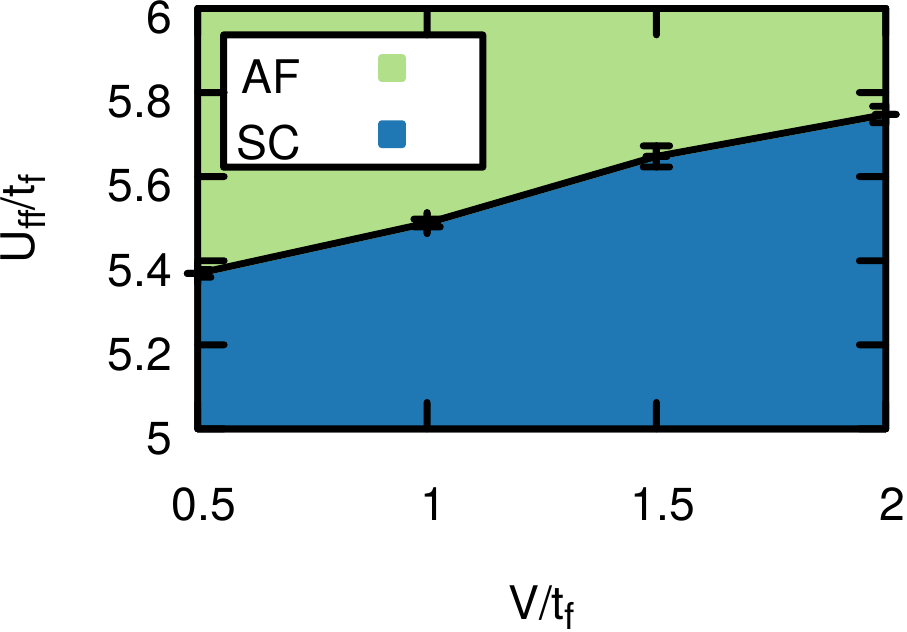}} 
\caption{ (a) Phase diagram based on the transition temperatures obtained from the linear extrapolation of the inverse pairing field susceptibilities $P_g$ for the competing instabilities towards $s$- and $d$-wave pairing for bosonic parameters $U_{\rm bb}/t_{\rm f}=5$ and $t_{\rm b}/t_{\rm f}=10$. 
The phase transition line corresponds to the critical coupling $U^{c}_{\rm ff}$ where the computed transition temperatures for the respective phases cross. The dashed lines correspond to the cuts along which the transition temperatures are shown in Fig.~\ref{fig:transition_tem_cuts}.
(b) Phase diagram focusing on the transition between the cluster antiferromagnetic instability ('AF') and the superconducting phase ('SC').
\label{fig:phase_diag}}
\end{figure}

Based on the extrapolated transition temperatures we obtain the phase-diagram in Fig~(\ref{fig:phase_diag})(a). For the bosonic parameters $U_{\rm bb}/t_{\rm f}=5$ and $t_{\rm b}/t_{\rm f}=10$ we have that $W \approx 0.25 V$ and we expect $s$-wave pairing to become relevant around $V \approx 4 W \sim 4 |U_{\rm ff}|$. Indeed, on increasing $U_{\rm bf}$ we first observe increasing transition temperatures for $d$-wave pairing up to the regime $|W| \gtrsim |U_{\rm ff}|$. 
\begin{figure} 
\centering 
\subfloat{\subfigimg[hsep=0.0\textwidth,vsep=0.0\textwidth,width=0.24\textwidth,pos=ul]{\bf (a)}{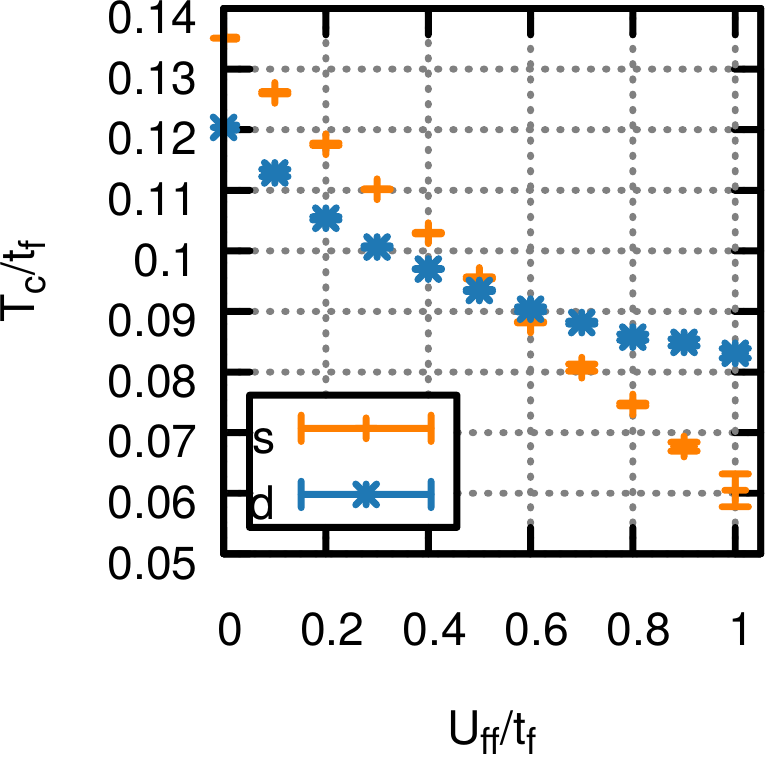}}
\hfill
\subfloat{\subfigimg[hsep=0.0\textwidth,vsep=0.0\textwidth,width=0.24\textwidth,pos=ul]{\bf (b)}{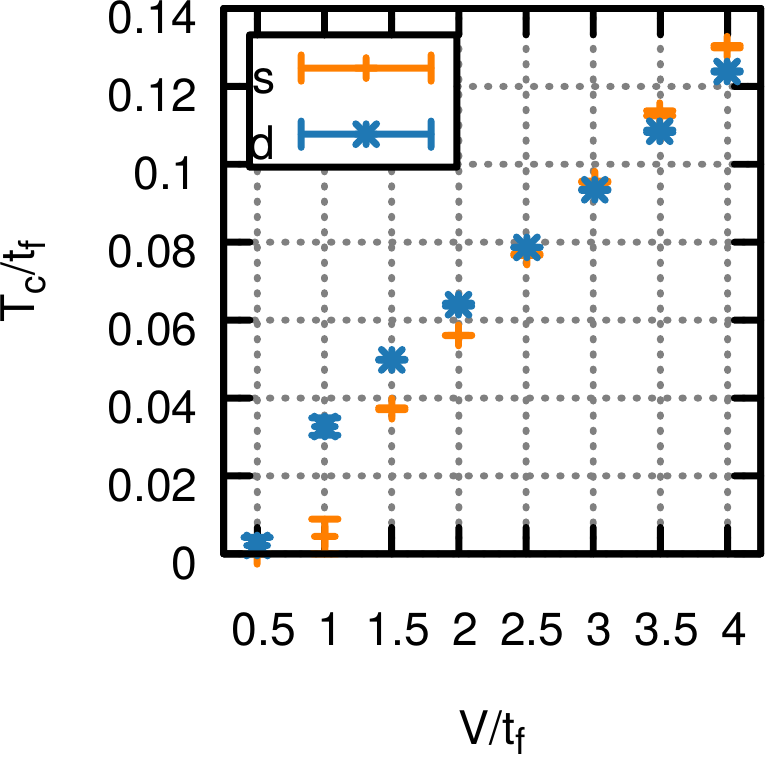}} 
\caption{Critical temperature $T_c$ for $s$- and $d$-wave pairing as a function of $U_{\rm ff}$ at fixed $V/t_{\rm f}=3$ in (a) and as a function of $V$ at fixed $U_{\rm ff}/t_{\rm f} =0.5 $ in (b).
\label{fig:transition_tem_cuts}}
\end{figure}

The critical temperatures for $s$- and $d$-wave pairing along cuts in the phase-diagram are shown in Fig~(\ref{fig:transition_tem_cuts}). As the $2\times2$ cluster represents the mean-field result for $d$-wave, underestimating the fluctuations and consequently overestimating the $d$-wave transition temperatures, we expect the transition line to shift in favor of $s$-wave in those regions in which the $s$-wave transition temperature is finite, {\it i.e.},  qualitatively those for which the effective induced interaction $W$ is stronger than the repulsion given by $U_{\rm ff}$.

Next we consider the instability towards an antiferromagnetic state. In the thermodynamical limit antiferromagnetism is allowed at zero temperature only but in selfconsistent cluster approaches such as DCA it can be found at finite temperature as well, rather accurately reproducing correlations at short distances but missing their fluctuations at long distances. 

\begin{figure} 
\centering 
\subfloat{\subfigimg[hsep=0.0\textwidth,vsep=0.0\textwidth,width=0.249\textwidth,pos=ul]{\bf (a)}{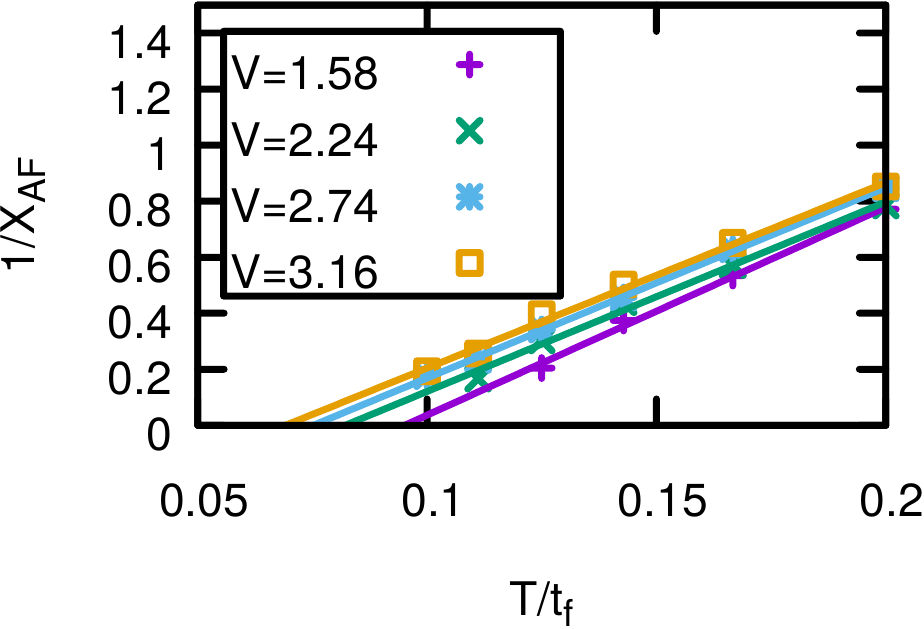}}
\subfloat{\subfigimg[hsep=0.0\textwidth,vsep=0.0\textwidth,width=0.249\textwidth,pos=ul]{\bf (b)}{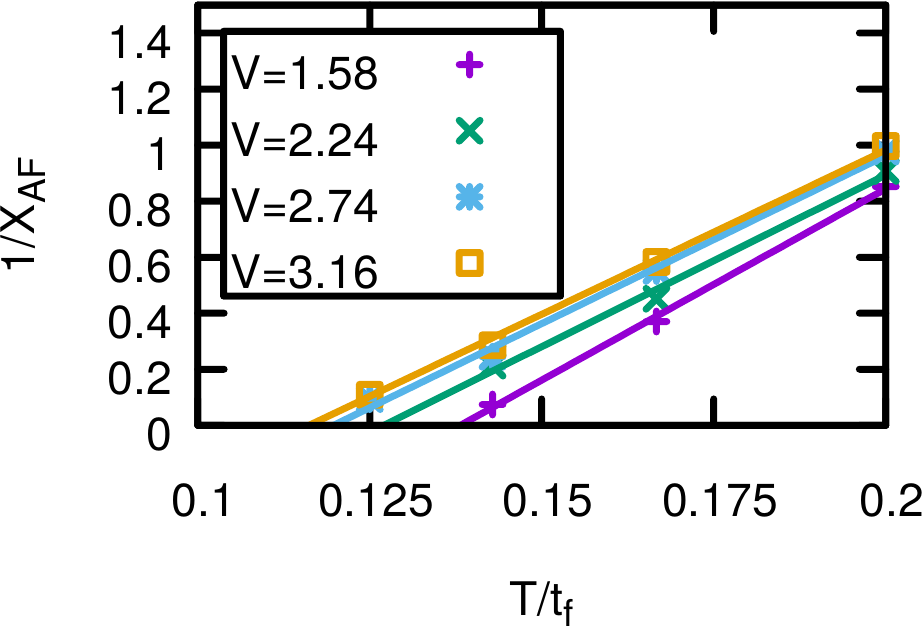}} 
  \caption{Inverse pairing field susceptibility $1/{\chi}_{AF}$ as a function of temperature $T/t_f$ at fixed $U_{\rm ff}= 5.2 (5.8)$ to the left (right) and fermionic half filling for different coupling strength $U_{\rm bf}$ to the bosons corresponding to $V=0.5,1.0,1.5,2.0$ for the bosonic parameters $U_{\rm bb}/t_{\rm f}=5$ and $t_{\rm b}/t_{\rm f}=10$. The vanishing of $1/P_{AF}$ signals the transition to an AFM state. The solid lines are linear fits to the lowest temperature points and are used to derive $T_c$.
\label{fig:pairfield_AF}}
\end{figure}

The spin susceptibility $\chi_{AF}(T)$, corresponding to an instability in the spin sector of the particle-hole channel, is shown in Fig.~\ref{fig:pairfield_AF} for $U_{\rm ff}= 5.2,5.8$ and for different values of the strength of the induced interaction $V=0.5,1.0,1.5,2.0$. To obtain the Neel temperature which corresponds to the divergence of $\chi_{AF}(T)$ a linear function to the inverse spin susceptibility in the low-temperature region is used as explained above. 

Based on the extrapolated transition temperatures for the divergence of the antiferromagnetic susceptibility $\chi_{\rm AF}$, we obtain a transition to an AFM state as shown in the corresponding phase diagram in Fig~(\ref{fig:phase_diag})(b). Again a stronger Fermi-Bose coupling $V$ tends to stabilise the $d$-wave state, in turn shifting the transition to the AFM state to higher values of $U_{\rm ff}$. The transition to the AFM state appears at considerably higher $U_{\rm ff}$ than in \cite{Mathey2006}, because the $2 \times 2$ cluster is known to strongly suppress antiferromagnetism \cite{Maier2005,Maier2005a}, thus favoring the superconducting state.

From these observations it emerges  that increasing $V$ enhances both $s$- and $d$-wave (with $s$-wave being dominant for nearly-free fermions). Increasing $U_{\rm ff}$ increases the spin density wave fluctuations and suppresses $s$-wave pairing, as a result of which $d$- can become dominant over an extended regime in parameter space. This suggests that $T_c$ for $d$-wave can be increased by increasing both $V$ and $U_{\rm ff}$ and staying near the line where $s$-wave becomes subdominant compared to $d$-wave. In the static mean-field study of Ref.~\cite{Wang2005} the highest $T_c$ for a $\mathrm{NaK}$ mixture was found at $\xi = 0.6$. For those parameters we find a maximal $T_c/t = 0.17$. Whereas Migdal-Eliashberg theories usually do not have a built-in mechanism that puts a limit on $T_c$,  the gain stops here when  $U_{\rm bf}$ is too large compared to $U_{\rm ff}$ and $U_{\rm bb}$ causing a mechanical instability towards phase-separation~\cite{Buchler_2003,Buchler_2004}.
%
\section{\label{sec:summary}Conclusions}
In this work we studied the antiferromagnetic und superconducting instabilities of Bose-Fermi-Mixtures in the limit where the bosons can be treated within the Bogoliubov approximation. We used a weak coupling Monte-Carlo cluster solver within the DCA-framework allowing us to distinguish between $s$- and $d$-wave pairing in the fermionic sector. The dominant pairing mechanism is determined by the relative sizes of the repulsive fermi-fermi interaction $U_{\rm ff}$ and the interactions induced by the bosons scaling with $V=U_{\rm bf}^{2}/ U_{\rm bb}$ leading to the phase diagram shown in Fig~(\ref{fig:phase_diag})(a). In particular, we find that $d$-wave superconductivity can be stabilized by the presence of bosonic particles. This extends the results of a recent DMFT study \cite{Anders2012} which found phases in which both bosons and fermions are superfluid, but could not determine the symmetry of the superfluid state.
In addition, for strong Fermi-Fermi interaction, we observe a transition to an antiferromagnetic state suppressing the superconducting pairing as seen in Fig~(\ref{fig:phase_diag})(b), consistent with a previous study based on the functional renormalization group \cite{Mathey2006}.
Our approach holds whenever the bosonic speed of sound $c$ is larger than the Fermi velocity $v_{\rm F}$. We checked that for $c/v_{\rm F} = 2$ the phase diagram looks similar but with stronger retardation effects. Our work can straightforwardly be extended to doped systems, allowing to address the strange metal physics and where also $p-$wave phases are predicted. The mechanisms discussed here are equally valid in 3d where $T_c$ for $d$- wave could well be much higher than in $2d$. \\

{\it Acknowledgments} -- 
We wish to thank M. Bukov and M. Punk for fruitful discussions and J. P. F. LeBlanc for providing testresults to benchmark our DCA code.  This work is supported by EPSRC Grant EP/K030094/1 and FP7/ERC Starting Grant No. 306897. Use was made of the ALPS libaries~\cite{ALPS,ALPS_DMFT}. LP thanks the hospitality of the Aspen Center for Physics (NSF Grant No. 1066293).

\appendix
\section{Derivation of the effective action}\label{app:eff_action_construction}
In this appendix we explicitly discuss the construction of the effective action in Eq.~\ref{eq:action}. Starting from the Hamiltonian of Bose-Fermi-Mixtures on a 2D square lattice
\begin{align}
H =& H_{\rm f} +H_{\rm b}+ H_{\rm bf} \notag  \\
  = &  - t_f \sum_{<i,j>, \sigma}{c_{i,\sigma}^{\dagger} c_{j,\sigma}} - \mu_f \sum_{i,\sigma}{ n_{i,\sigma}^{f}} + U_{\rm ff} \sum_i{n_{i,\uparrow}^{f} n_{i,\downarrow}^{f}} \notag \\
         &  - t_b\, \sum_{<i,j>}{b_i^{\dagger} \, b_j} - \mu_b\sum{n_i^{b}} + \frac{U_{\rm bb}}{2} \sum{n_i^{b}(n_i^{b}-1)}  \\
         &  + U_{\rm bf} \sum_{i,\sigma}{n_{i}^{b} \, n_{i,\sigma}^{f}} \notag
\end{align}
where $c_i^{\dagger}$ ($b_i^{\dagger})$ are fermionic (bosonic) creation operators at site $i$, $n_i^{f(b)}$ the corresponding densities, $t_{\rm f(b)}$ describes the hopping of a fermion (boson) from site $i$ to site $j$, $\mu_{\rm f(b)}$ is the chemical potential for fermions (bosons), $U_{\rm ff(bb)}$ is the on-site repulsion of fermions (bosons) and $U_{\rm bf}$ the on-site interaction between bosons and fermions. The model is the sum of a Fermi-Hubbard model (first line) the Bose-Hubbard model (second line) and a density-density interaction between bosons and fermions (third line). In the following we will treat the bosons within the Bogoliubov approximation allowing us to integrate them out in favour of a retarded density-density interaction between fermions.

In the Bogoliubov approximation the bosonic Hamiltonian $H_b$ takes the form $H_b \approx \sum_{\vect{k}} E_{\vect{k}} \, \alpha_{\vect{k}}^{\dagger} \alpha_{\vect{k}} $ in terms of the Bogoliubov quasi-particles defined by $b_{\vect{k}} =u_{\vect{k}} \alpha_{\vect{k}} - v_{\vect{k}} \alpha_{-\vect{k}}^{\dagger} $ and $ b^{\dagger}_{\vect{k}} =u_{\vect{k}} \alpha^{\dagger}_{\vect{k}} - v_{\vect{k}} \alpha_{-\vect{k}}$ respectively. As usual the coefficients are given by $ u_{\vect{k}} = \cosh{\phi_{\vect{k}}}$ and $v_{\vect{k}} = \sinh{\phi_{\vect{k}}}$ where  $\tanh{2 \, \phi_{\vect{k}}} = n_0 U_{\rm bb}/(\bar{\varepsilon}_{\vect{k}} + n_0 U_{\rm bb})$.
The Bogoliubov spectrum is given by $E_{\vect{k}}=\left[\bar{\varepsilon}_{\vect{k}}^{2}+ 2 \bar{\varepsilon}_{\vect{k}} n_0 U_{\rm bb} \right]^{1/2}$ with $\bar{\varepsilon}_{\vect{k}} = \varepsilon_{\vect{k}} - \varepsilon_{\vect{0}}$, the lattice dispersion of non-interacting bosons shifted to be positive.
We emphasize that this treatment neglects any backaction of the fermionic sector on the bosons, in particular we assume that the bosons do not charge order, which limits our approach to the case of fast bosons.

Rewriting the coupling term $H_{\rm bf}$ in momentum space we obtain
\begin{align}
 H_{\rm bf}&= \frac{1}{N_s}\sum_{i,\vec{k}_1, \vect{k}_2}{ e^{-i(\vect{k}_1-\vect{k}_2)\vect{r}_i} b_{\vect{k}_1}^{\dagger} b_{\vect{k}_2} \, n_{i}^{f}} \\
                                                &= \frac{N_0}{N_s} \sum_{i}{ n_{i}^{f} } + \frac{\sqrt{N_0}}{N_s} \sum^{\prime}_{i, k}{\left(e^{i k r_i} b_{\vect{k}} +e^{-i k r_i} b_{\vect{k}}^{\dagger} \right) \, n_{i}^{f}} \notag \\
                                                &\qquad  + \frac{1}{N_s}\sum^{\prime}_{i, \vect{k}_1, \vect{k}_2}{ e^{-i(\vect{k}_1-\vect{k}_2)r_i} b_{\vect{k}_1}^{\dagger} b_{\vect{k}_2} \, n_{i}^{f}}
\end{align}
where $N_0$ is the macroscopic number of bosonic atoms in the condensate, $N_s$ is the number of sites in the lattice and the primed sum indicates that the $k=0$ terms should be omitted. In the following we will neglect the last term as it is suppressed by a power of $\sqrt{N_0}$ compared to the second term. The first term corresponds to the shift in the fermionic chemical potential present in Eq.~\ref{eq:action} whereas the second term represents the linear coupling of bosons to the fermions that will be integrated out.

Next, we reexpress the second term in the Bogoliubov boson operators to obtain
\begin{multline}
   \sum^{\prime}_{i \vect{k}}{\left(e^{i \vect{k} \vect{r}_i} b_{\vect{k}} +e^{-i \vect{k} \vect{r}_i} b_{\vect{k}}^{\dagger} \right) \, n_{i}^{f}}\\
  =  \sum^{\prime}_{i, \vect{k}}{\left(        \alpha_{\vect{k}} \left[u_{\vect{k}} - v_{-\vect{k}}  \right] e^{i \vect{k} \vect{ r}_i} + \alpha_{\vect{k}}^{\dagger} \left[u_{\vect{k}} - v_{-\vect{k}}  \right] e^{-i \vect{k} \vect{r}_i} \right) \, n_{i}^{f}   } \, .
\end{multline}
which clearly shows that the Bogoliubov bosons couple linearly to the fermionic density with momentum dependent couplings.

Employing coherent states for both the bosons and fermions the action reads as $S = S_{\rm f} +S_{\rm bog} +S_{\rm bf} $ where the fermionic action $S_f$ has already been defined in Eq.~\ref{eq:action} and
\begin{align}
S_{\rm bog} &= \int_0^{\beta} {d\tau \, \sum_{\vect{k}}{\bar{\alpha}_{\vect{k}}(\tau)\left[ E_{\vect{k}} +\frac{\partial}{\partial \tau}  \right]}  \alpha_{\vect{k}}(\tau)} \\
S_{\rm bf}  &= \sqrt{n_0} \, U_{\rm bf} \int_0^{\beta} d\tau \,\frac{1}{\sqrt{N_s}} \sum_{i, \vect{k}} \Big(     \alpha_{\vect{k}}(\tau) \left[u_{\vect{k}} - v_{-\vect{k}}  \right] e^{i \vect{k} \vect{r}_i} \notag  \\
        & \qquad   + \bar{\alpha}_{\vect{k}}(\tau) \left[u_{\vect{k}} - v_{-\vect{k}}  \right] e^{-i \vect{k} \vect{r}_i}  \Big) n_{i}^{f} (\tau) 
\end{align}

Finally, integration over the quadratic bosonic Bogoliubov action yields the non-local, retarded density-density interaction, i.e.
$ e^{-S_{ret}} =\int{d(\bar{\alpha},\alpha)\, e^{-S_{\rm bog} -S_{\rm bf}}} $ with 
\begin{equation}
 S_{\rm ret} = - \frac{n_0 \, U^{2}_{\rm bf}}{2} \iint_0^{\beta}{d\tau_1 d\tau_2 \,\sum_{i,j}}n_{i}^{f}(\tau_1)\, D_{ij}(\tau_1-\tau_2)\, n_{j}^{f}(\tau_2)
\end{equation}
which combined with $S_f$ gives the total action $S=S_f+S_{\rm ret}$ of the effective model in Eq.~\ref{eq:action} and the kernel $D_{ij}$ has been defined in Eq.~\ref{eq:ret_int_kernel}.

\section{Discussion of the sign problem}\label{app:sign}
\begin{figure} 
\centering 
\subfloat{\subfigimg[hsep=0.0\textwidth,vsep=0.0\textwidth,width=0.24\textwidth,pos=ul]{\bf (a)}{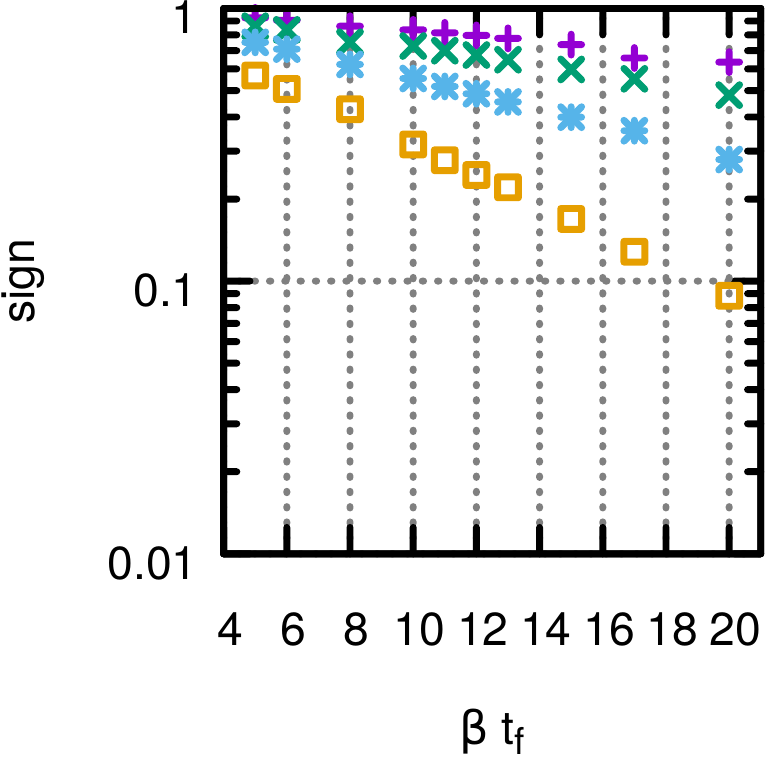}}
\subfloat{\subfigimg[hsep=0.02\textwidth,vsep=0.0\textwidth,width=0.24\textwidth,pos=ul]{\bf (b)}{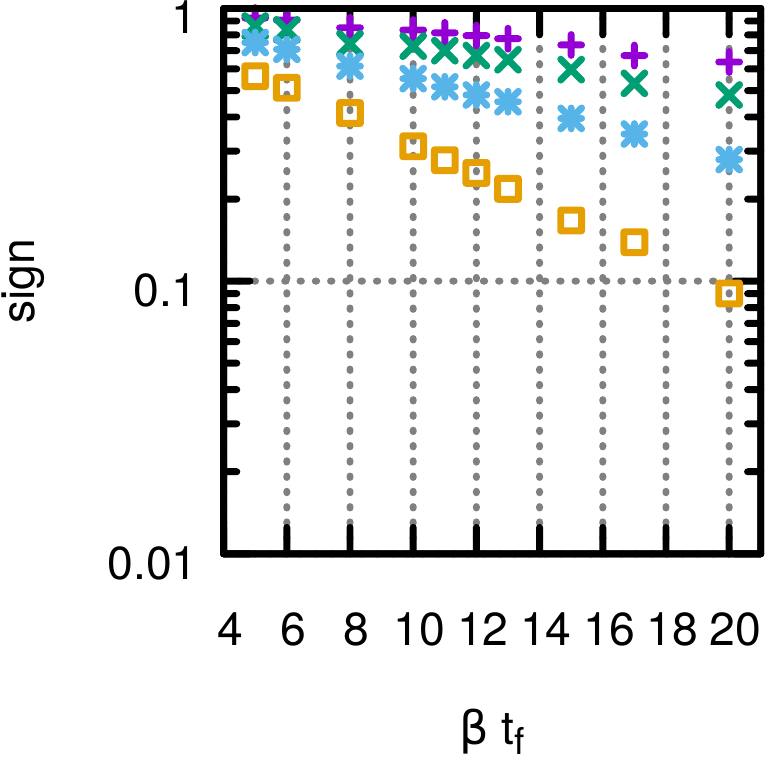}} 

\subfloat{\subfigimg[hsep=0.0\textwidth,vsep=0.0\textwidth,width=0.24\textwidth,pos=ul]{\bf (c)}{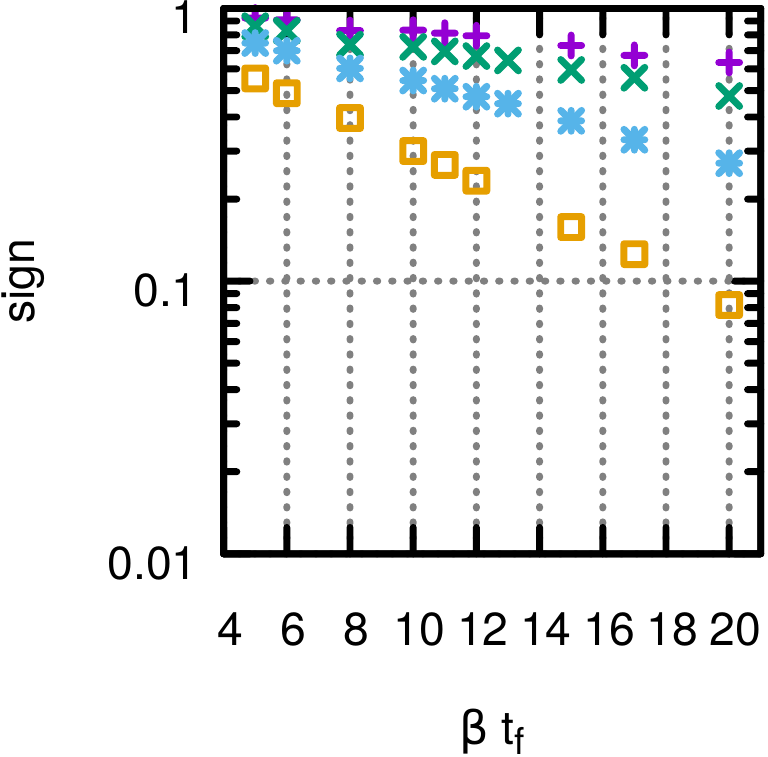}}
\subfloat{\subfigimg[hsep=0.02\textwidth,vsep=0.0\textwidth,width=0.24\textwidth,pos=ul]{\bf (d)}{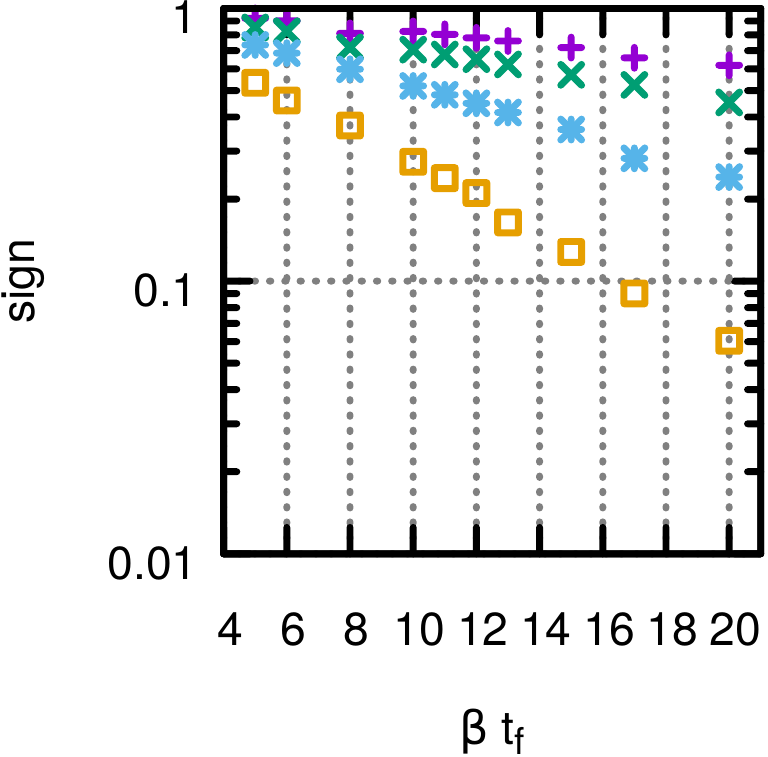}} 
  \caption{Average sign of the Monte-Carlo simulations as a function of $\beta$ for $U_{\rm ff}/t_{f} = 1,2,3,4$ in panels (a)-(d), in each case for $V/t_f=0.5,1.0,2.0,4.0$ from top to bottom.
\label{fig:app_sign}}
\end{figure}

As we are at half-filling the simulations are sign-free for $V=0$. However, as discussed in the main text the induced interactions lead to a sign-problem. The average sign of the simulations is shown in Fig.~\ref{fig:app_sign} in each panel at a fixed value of $U_{\rm ff}$ for different values of $V=U_{\rm bf}^2/U_{\rm bb}$.

We only observe a very weak dependence of the sign on the fermi-fermi interaction $U_{\rm ff }$ as expected from the sign free character for $V=0$.
In all cases the sign is exponentially decreasing as a function of $\beta$ with a decay rate that increases as $V$ is increased and the induced interactions become more important.
This exponential dependence on the temperatures limits the lowest temperatures we can access in the simulations. Moreover, the strong dependence on the induced interactions limits the maximally accessible $V$ and ultimately the extension to bigger clusters where the number of interaction terms proliferates.

\bibliography{BFM_database_new}{}

\begin{thebibliography}{81}%
\makeatletter
\providecommand \@ifxundefined [1]{%
 \@ifx{#1\undefined}
}%
\providecommand \@ifnum [1]{%
 \ifnum #1\expandafter \@firstoftwo
 \else \expandafter \@secondoftwo
 \fi
}%
\providecommand \@ifx [1]{%
 \ifx #1\expandafter \@firstoftwo
 \else \expandafter \@secondoftwo
 \fi
}%
\providecommand \natexlab [1]{#1}%
\providecommand \enquote  [1]{``#1''}%
\providecommand \bibnamefont  [1]{#1}%
\providecommand \bibfnamefont [1]{#1}%
\providecommand \citenamefont [1]{#1}%
\providecommand \href@noop [0]{\@secondoftwo}%
\providecommand \href [0]{\begingroup \@sanitize@url \@href}%
\providecommand \@href[1]{\@@startlink{#1}\@@href}%
\providecommand \@@href[1]{\endgroup#1\@@endlink}%
\providecommand \@sanitize@url [0]{\catcode `\\12\catcode `\$12\catcode
  `\&12\catcode `\#12\catcode `\^12\catcode `\_12\catcode `\%12\relax}%
\providecommand \@@startlink[1]{}%
\providecommand \@@endlink[0]{}%
\providecommand \url  [0]{\begingroup\@sanitize@url \@url }%
\providecommand \@url [1]{\endgroup\@href {#1}{\urlprefix }}%
\providecommand \urlprefix  [0]{URL }%
\providecommand \Eprint [0]{\href }%
\providecommand \doibase [0]{http://dx.doi.org/}%
\providecommand \selectlanguage [0]{\@gobble}%
\providecommand \bibinfo  [0]{\@secondoftwo}%
\providecommand \bibfield  [0]{\@secondoftwo}%
\providecommand \translation [1]{[#1]}%
\providecommand \BibitemOpen [0]{}%
\providecommand \bibitemStop [0]{}%
\providecommand \bibitemNoStop [0]{.\EOS\space}%
\providecommand \EOS [0]{\spacefactor3000\relax}%
\providecommand \BibitemShut  [1]{\csname bibitem#1\endcsname}%
\let\auto@bib@innerbib\@empty
\bibitem [{\citenamefont {Bloch}\ and\ \citenamefont
  {Zwerger}(2008)}]{Bloch2008}%
  \BibitemOpen
  \bibfield  {author} {\bibinfo {author} {\bibfnamefont {I.}~\bibnamefont
  {Bloch}}\ and\ \bibinfo {author} {\bibfnamefont {W.}~\bibnamefont
  {Zwerger}},\ }\href {\doibase 10.1103/revmodphys.80.885} {\bibfield
  {journal} {\bibinfo  {journal} {Rev. Mod. Phys.}\ }\textbf {\bibinfo {volume}
  {80}},\ \bibinfo {pages} {885–964} (\bibinfo {year} {2008})}\BibitemShut
  {NoStop}%
\bibitem [{\citenamefont {{Ketterle}}\ and\ \citenamefont
  {{Zwierlein}}(2008)}]{Ketterle_2008}%
  \BibitemOpen
  \bibfield  {author} {\bibinfo {author} {\bibfnamefont {W.}~\bibnamefont
  {{Ketterle}}}\ and\ \bibinfo {author} {\bibfnamefont {M.~W.}\ \bibnamefont
  {{Zwierlein}}},\ }\href {\doibase 10.1393/ncr/i2008-10033-1} {\bibfield
  {journal} {\bibinfo  {journal} {Nuovo Cimento Rivista Serie}\ }\textbf
  {\bibinfo {volume} {31}},\ \bibinfo {pages} {247} (\bibinfo {year} {2008})},\
  \Eprint {http://arxiv.org/abs/0801.2500} {arXiv:0801.2500 [cond-mat.other]}
  \BibitemShut {NoStop}%
\bibitem [{\citenamefont {Wang}\ \emph {et~al.}(2005)\citenamefont {Wang},
  \citenamefont {Lukin},\ and\ \citenamefont {Demler}}]{Wang2005}%
  \BibitemOpen
  \bibfield  {author} {\bibinfo {author} {\bibfnamefont {D.-W.}\ \bibnamefont
  {Wang}}, \bibinfo {author} {\bibfnamefont {M.~D.}\ \bibnamefont {Lukin}}, \
  and\ \bibinfo {author} {\bibfnamefont {E.}~\bibnamefont {Demler}},\ }\href
  {\doibase 10.1103/PhysRevA.72.051604} {\bibfield  {journal} {\bibinfo
  {journal} {Phys. Rev. A}\ }\textbf {\bibinfo {volume} {72}},\ \bibinfo
  {pages} {051604} (\bibinfo {year} {2005})}\BibitemShut {NoStop}%
\bibitem [{\citenamefont {Scalapino}\ \emph {et~al.}(1986)\citenamefont
  {Scalapino}, \citenamefont {Loh},\ and\ \citenamefont
  {Hirsch}}]{Scalapino_1986}%
  \BibitemOpen
  \bibfield  {author} {\bibinfo {author} {\bibfnamefont {D.~J.}\ \bibnamefont
  {Scalapino}}, \bibinfo {author} {\bibfnamefont {E.}~\bibnamefont {Loh}}, \
  and\ \bibinfo {author} {\bibfnamefont {J.~E.}\ \bibnamefont {Hirsch}},\
  }\href {\doibase 10.1103/PhysRevB.34.8190} {\bibfield  {journal} {\bibinfo
  {journal} {Phys. Rev. B}\ }\textbf {\bibinfo {volume} {34}},\ \bibinfo
  {pages} {8190} (\bibinfo {year} {1986})}\BibitemShut {NoStop}%
\bibitem [{\citenamefont {Abanov}\ \emph {et~al.}(2001)\citenamefont {Abanov},
  \citenamefont {Chubukov},\ and\ \citenamefont {Finkel'stein}}]{Abanov_2001}%
  \BibitemOpen
  \bibfield  {author} {\bibinfo {author} {\bibfnamefont {A.}~\bibnamefont
  {Abanov}}, \bibinfo {author} {\bibfnamefont {A.~V.}\ \bibnamefont
  {Chubukov}}, \ and\ \bibinfo {author} {\bibfnamefont {A.~M.}\ \bibnamefont
  {Finkel'stein}},\ }\href {http://stacks.iop.org/0295-5075/54/i=4/a=488}
  {\bibfield  {journal} {\bibinfo  {journal} {EPL (Europhysics Letters)}\
  }\textbf {\bibinfo {volume} {54}},\ \bibinfo {pages} {488} (\bibinfo {year}
  {2001})}\BibitemShut {NoStop}%
\bibitem [{\citenamefont {Abanov}\ \emph {et~al.}(2003)\citenamefont {Abanov},
  \citenamefont {Chubukov},\ and\ \citenamefont {Schmalian}}]{Abanov_2003}%
  \BibitemOpen
  \bibfield  {author} {\bibinfo {author} {\bibfnamefont {A.}~\bibnamefont
  {Abanov}}, \bibinfo {author} {\bibfnamefont {A.~V.}\ \bibnamefont
  {Chubukov}}, \ and\ \bibinfo {author} {\bibfnamefont {J.}~\bibnamefont
  {Schmalian}},\ }\href {\doibase 10.1080/0001873021000057123} {\bibfield
  {journal} {\bibinfo  {journal} {Advances in Physics}\ }\textbf {\bibinfo
  {volume} {52}},\ \bibinfo {pages} {119–218} (\bibinfo {year} {2003})},\
  \Eprint {http://arxiv.org/abs/http://dx.doi.org/10.1080/0001873021000057123}
  {http://dx.doi.org/10.1080/0001873021000057123} \BibitemShut {NoStop}%
\bibitem [{\citenamefont {Wang}\ and\ \citenamefont
  {Chubukov}(2013)}]{Chubukov_2013}%
  \BibitemOpen
  \bibfield  {author} {\bibinfo {author} {\bibfnamefont {Y.}~\bibnamefont
  {Wang}}\ and\ \bibinfo {author} {\bibfnamefont {A.~V.}\ \bibnamefont
  {Chubukov}},\ }\href {\doibase 10.1103/PhysRevLett.110.127001} {\bibfield
  {journal} {\bibinfo  {journal} {Phys. Rev. Lett.}\ }\textbf {\bibinfo
  {volume} {110}},\ \bibinfo {pages} {127001} (\bibinfo {year}
  {2013})}\BibitemShut {NoStop}%
\bibitem [{\citenamefont {Metlitski}\ and\ \citenamefont
  {Sachdev}(2010{\natexlab{a}})}]{Metlitski_2010a}%
  \BibitemOpen
  \bibfield  {author} {\bibinfo {author} {\bibfnamefont {M.~A.}\ \bibnamefont
  {Metlitski}}\ and\ \bibinfo {author} {\bibfnamefont {S.}~\bibnamefont
  {Sachdev}},\ }\href {http://stacks.iop.org/1367-2630/12/i=10/a=105007}
  {\bibfield  {journal} {\bibinfo  {journal} {New Journal of Physics}\ }\textbf
  {\bibinfo {volume} {12}},\ \bibinfo {pages} {105007} (\bibinfo {year}
  {2010}{\natexlab{a}})}\BibitemShut {NoStop}%
\bibitem [{\citenamefont {Metlitski}\ and\ \citenamefont
  {Sachdev}(2010{\natexlab{b}})}]{Metlitski_2010b}%
  \BibitemOpen
  \bibfield  {author} {\bibinfo {author} {\bibfnamefont {M.~A.}\ \bibnamefont
  {Metlitski}}\ and\ \bibinfo {author} {\bibfnamefont {S.}~\bibnamefont
  {Sachdev}},\ }\href {\doibase 10.1103/PhysRevB.82.075128} {\bibfield
  {journal} {\bibinfo  {journal} {Phys. Rev. B}\ }\textbf {\bibinfo {volume}
  {82}},\ \bibinfo {pages} {075128} (\bibinfo {year}
  {2010}{\natexlab{b}})}\BibitemShut {NoStop}%
\bibitem [{\citenamefont {Berg}\ \emph {et~al.}(2012)\citenamefont {Berg},
  \citenamefont {Metlitski},\ and\ \citenamefont {Sachdev}}]{Berg_2011}%
  \BibitemOpen
  \bibfield  {author} {\bibinfo {author} {\bibfnamefont {E.}~\bibnamefont
  {Berg}}, \bibinfo {author} {\bibfnamefont {M.~A.}\ \bibnamefont {Metlitski}},
  \ and\ \bibinfo {author} {\bibfnamefont {S.}~\bibnamefont {Sachdev}},\ }\href
  {\doibase 10.1126/science.1227769} {\bibfield  {journal} {\bibinfo  {journal}
  {Science}\ }\textbf {\bibinfo {volume} {338}},\ \bibinfo {pages} {1606}
  (\bibinfo {year} {2012})},\ \Eprint
  {http://arxiv.org/abs/http://www.sciencemag.org/content/338/6114/1606.full.pdf}
  {http://www.sciencemag.org/content/338/6114/1606.full.pdf} \BibitemShut
  {NoStop}%
\bibitem [{\citenamefont {Moon}\ and\ \citenamefont
  {Sachdev}(2009)}]{Sachdev_2009}%
  \BibitemOpen
  \bibfield  {author} {\bibinfo {author} {\bibfnamefont {E.~G.}\ \bibnamefont
  {Moon}}\ and\ \bibinfo {author} {\bibfnamefont {S.}~\bibnamefont {Sachdev}},\
  }\href {\doibase 10.1103/PhysRevB.80.035117} {\bibfield  {journal} {\bibinfo
  {journal} {Phys. Rev. B}\ }\textbf {\bibinfo {volume} {80}},\ \bibinfo
  {pages} {035117} (\bibinfo {year} {2009})}\BibitemShut {NoStop}%
\bibitem [{\citenamefont {Sachdev}\ \emph {et~al.}(2012)\citenamefont
  {Sachdev}, \citenamefont {Metlitski},\ and\ \citenamefont
  {Punk}}]{Sachdev_2012}%
  \BibitemOpen
  \bibfield  {author} {\bibinfo {author} {\bibfnamefont {S.}~\bibnamefont
  {Sachdev}}, \bibinfo {author} {\bibfnamefont {M.~A.}\ \bibnamefont
  {Metlitski}}, \ and\ \bibinfo {author} {\bibfnamefont {M.}~\bibnamefont
  {Punk}},\ }\href {http://stacks.iop.org/0953-8984/24/i=29/a=294205}
  {\bibfield  {journal} {\bibinfo  {journal} {Journal of Physics: Condensed
  Matter}\ }\textbf {\bibinfo {volume} {24}},\ \bibinfo {pages} {294205}
  (\bibinfo {year} {2012})}\BibitemShut {NoStop}%
\bibitem [{\citenamefont {Scalapino}(2012)}]{Scalapino_2012}%
  \BibitemOpen
  \bibfield  {author} {\bibinfo {author} {\bibfnamefont {D.~J.}\ \bibnamefont
  {Scalapino}},\ }\href {\doibase 10.1103/RevModPhys.84.1383} {\bibfield
  {journal} {\bibinfo  {journal} {Rev. Mod. Phys.}\ }\textbf {\bibinfo {volume}
  {84}},\ \bibinfo {pages} {1383} (\bibinfo {year} {2012})}\BibitemShut
  {NoStop}%
\bibitem [{\citenamefont {Zhai}\ \emph {et~al.}(2009)\citenamefont {Zhai},
  \citenamefont {Wang},\ and\ \citenamefont {Lee}}]{Zhai_2009}%
  \BibitemOpen
  \bibfield  {author} {\bibinfo {author} {\bibfnamefont {H.}~\bibnamefont
  {Zhai}}, \bibinfo {author} {\bibfnamefont {F.}~\bibnamefont {Wang}}, \ and\
  \bibinfo {author} {\bibfnamefont {D.-H.}\ \bibnamefont {Lee}},\ }\href
  {\doibase 10.1103/PhysRevB.80.064517} {\bibfield  {journal} {\bibinfo
  {journal} {Phys. Rev. B}\ }\textbf {\bibinfo {volume} {80}},\ \bibinfo
  {pages} {064517} (\bibinfo {year} {2009})}\BibitemShut {NoStop}%
\bibitem [{\citenamefont {Moon}\ and\ \citenamefont
  {Sachdev}(2010)}]{Moon_2010}%
  \BibitemOpen
  \bibfield  {author} {\bibinfo {author} {\bibfnamefont {E.~G.}\ \bibnamefont
  {Moon}}\ and\ \bibinfo {author} {\bibfnamefont {S.}~\bibnamefont {Sachdev}},\
  }\href {\doibase 10.1103/PhysRevB.82.104516} {\bibfield  {journal} {\bibinfo
  {journal} {Phys. Rev. B}\ }\textbf {\bibinfo {volume} {82}},\ \bibinfo
  {pages} {104516} (\bibinfo {year} {2010})}\BibitemShut {NoStop}%
\bibitem [{\citenamefont {Kuroki}\ \emph {et~al.}(2008)\citenamefont {Kuroki},
  \citenamefont {Onari}, \citenamefont {Arita}, \citenamefont {Usui},
  \citenamefont {Tanaka}, \citenamefont {Kontani},\ and\ \citenamefont
  {Aoki}}]{Kuroki_2008}%
  \BibitemOpen
  \bibfield  {author} {\bibinfo {author} {\bibfnamefont {K.}~\bibnamefont
  {Kuroki}}, \bibinfo {author} {\bibfnamefont {S.}~\bibnamefont {Onari}},
  \bibinfo {author} {\bibfnamefont {R.}~\bibnamefont {Arita}}, \bibinfo
  {author} {\bibfnamefont {H.}~\bibnamefont {Usui}}, \bibinfo {author}
  {\bibfnamefont {Y.}~\bibnamefont {Tanaka}}, \bibinfo {author} {\bibfnamefont
  {H.}~\bibnamefont {Kontani}}, \ and\ \bibinfo {author} {\bibfnamefont
  {H.}~\bibnamefont {Aoki}},\ }\href {\doibase 10.1103/PhysRevLett.101.087004}
  {\bibfield  {journal} {\bibinfo  {journal} {Phys. Rev. Lett.}\ }\textbf
  {\bibinfo {volume} {101}},\ \bibinfo {pages} {087004} (\bibinfo {year}
  {2008})}\BibitemShut {NoStop}%
\bibitem [{\citenamefont {Mazin}\ \emph {et~al.}(2008)\citenamefont {Mazin},
  \citenamefont {Singh}, \citenamefont {Johannes},\ and\ \citenamefont
  {Du}}]{Mazin_2008}%
  \BibitemOpen
  \bibfield  {author} {\bibinfo {author} {\bibfnamefont {I.~I.}\ \bibnamefont
  {Mazin}}, \bibinfo {author} {\bibfnamefont {D.~J.}\ \bibnamefont {Singh}},
  \bibinfo {author} {\bibfnamefont {M.~D.}\ \bibnamefont {Johannes}}, \ and\
  \bibinfo {author} {\bibfnamefont {M.~H.}\ \bibnamefont {Du}},\ }\href
  {\doibase 10.1103/PhysRevLett.101.057003} {\bibfield  {journal} {\bibinfo
  {journal} {Phys. Rev. Lett.}\ }\textbf {\bibinfo {volume} {101}},\ \bibinfo
  {pages} {057003} (\bibinfo {year} {2008})}\BibitemShut {NoStop}%
\bibitem [{\citenamefont {Graser}\ \emph {et~al.}(2009)\citenamefont {Graser},
  \citenamefont {Maier}, \citenamefont {Hirschfeld},\ and\ \citenamefont
  {Scalapino}}]{Scalapino_2009}%
  \BibitemOpen
  \bibfield  {author} {\bibinfo {author} {\bibfnamefont {S.}~\bibnamefont
  {Graser}}, \bibinfo {author} {\bibfnamefont {T.~A.}\ \bibnamefont {Maier}},
  \bibinfo {author} {\bibfnamefont {P.~J.}\ \bibnamefont {Hirschfeld}}, \ and\
  \bibinfo {author} {\bibfnamefont {D.~J.}\ \bibnamefont {Scalapino}},\ }\href
  {http://stacks.iop.org/1367-2630/11/i=2/a=025016} {\bibfield  {journal}
  {\bibinfo  {journal} {New Journal of Physics}\ }\textbf {\bibinfo {volume}
  {11}},\ \bibinfo {pages} {025016} (\bibinfo {year} {2009})}\BibitemShut
  {NoStop}%
\bibitem [{\citenamefont {Maiti}\ and\ \citenamefont
  {Chubukov}(2010)}]{Chubukov_2010}%
  \BibitemOpen
  \bibfield  {author} {\bibinfo {author} {\bibfnamefont {S.}~\bibnamefont
  {Maiti}}\ and\ \bibinfo {author} {\bibfnamefont {A.~V.}\ \bibnamefont
  {Chubukov}},\ }\href {\doibase 10.1103/PhysRevB.82.214515} {\bibfield
  {journal} {\bibinfo  {journal} {Phys. Rev. B}\ }\textbf {\bibinfo {volume}
  {82}},\ \bibinfo {pages} {214515} (\bibinfo {year} {2010})}\BibitemShut
  {NoStop}%
\bibitem [{\citenamefont {{Paglione Johnpierre}}\ and\ \citenamefont {{Greene
  Richard L.}}(2010)}]{Paglione_2010}%
  \BibitemOpen
  \bibfield  {author} {\bibinfo {author} {\bibnamefont {{Paglione
  Johnpierre}}}\ and\ \bibinfo {author} {\bibnamefont {{Greene Richard L.}}},\
  }\href {\doibase http://dx.doi.org/10.1038/nphys1759} {\bibfield  {journal}
  {\bibinfo  {journal} {Nat Phys}\ }\textbf {\bibinfo {volume} {6}},\ \bibinfo
  {pages} {645–658} (\bibinfo {year} {2010})},\ \bibinfo {note}
  {10.1038/nphys1759}\BibitemShut {NoStop}%
\bibitem [{\citenamefont {Johnston}(2010)}]{Johnston_2010}%
  \BibitemOpen
  \bibfield  {author} {\bibinfo {author} {\bibfnamefont {D.~C.}\ \bibnamefont
  {Johnston}},\ }\href {\doibase 10.1080/00018732.2010.513480} {\bibfield
  {journal} {\bibinfo  {journal} {Advances in Physics}\ }\textbf {\bibinfo
  {volume} {59}},\ \bibinfo {pages} {803–1061} (\bibinfo {year} {2010})},\
  \Eprint {http://arxiv.org/abs/http://dx.doi.org/10.1080/00018732.2010.513480}
  {http://dx.doi.org/10.1080/00018732.2010.513480} \BibitemShut {NoStop}%
\bibitem [{\citenamefont {{Gegenwart Philipp}}\ \emph
  {et~al.}(2008)\citenamefont {{Gegenwart Philipp}}, \citenamefont {{Si
  Qimiao}},\ and\ \citenamefont {{Steglich Frank}}}]{Gegenwart_2008}%
  \BibitemOpen
  \bibfield  {author} {\bibinfo {author} {\bibnamefont {{Gegenwart Philipp}}},
  \bibinfo {author} {\bibnamefont {{Si Qimiao}}}, \ and\ \bibinfo {author}
  {\bibnamefont {{Steglich Frank}}},\ }\href {\doibase
  http://dx.doi.org/10.1038/nphys892} {\bibfield  {journal} {\bibinfo
  {journal} {Nat Phys}\ }\textbf {\bibinfo {volume} {4}},\ \bibinfo {pages}
  {186–197} (\bibinfo {year} {2008})},\ \bibinfo {note}
  {10.1038/nphys892}\BibitemShut {NoStop}%
\bibitem [{\citenamefont {Nair}\ \emph {et~al.}(2010)\citenamefont {Nair},
  \citenamefont {Stockert}, \citenamefont {Witte}, \citenamefont {Nicklas},
  \citenamefont {Schedler}, \citenamefont {Kiefer}, \citenamefont {Thompson},
  \citenamefont {Bianchi}, \citenamefont {Fisk}, \citenamefont {Wirth},\ and\
  \citenamefont {Steglich}}]{Nair_2010}%
  \BibitemOpen
  \bibfield  {author} {\bibinfo {author} {\bibfnamefont {S.}~\bibnamefont
  {Nair}}, \bibinfo {author} {\bibfnamefont {O.}~\bibnamefont {Stockert}},
  \bibinfo {author} {\bibfnamefont {U.}~\bibnamefont {Witte}}, \bibinfo
  {author} {\bibfnamefont {M.}~\bibnamefont {Nicklas}}, \bibinfo {author}
  {\bibfnamefont {R.}~\bibnamefont {Schedler}}, \bibinfo {author}
  {\bibfnamefont {K.}~\bibnamefont {Kiefer}}, \bibinfo {author} {\bibfnamefont
  {J.~D.}\ \bibnamefont {Thompson}}, \bibinfo {author} {\bibfnamefont {A.~D.}\
  \bibnamefont {Bianchi}}, \bibinfo {author} {\bibfnamefont {Z.}~\bibnamefont
  {Fisk}}, \bibinfo {author} {\bibfnamefont {S.}~\bibnamefont {Wirth}}, \ and\
  \bibinfo {author} {\bibfnamefont {F.}~\bibnamefont {Steglich}},\ }\href
  {\doibase 10.1073/pnas.1004958107} {\bibfield  {journal} {\bibinfo  {journal}
  {Proceedings of the National Academy of Sciences}\ }\textbf {\bibinfo
  {volume} {107}},\ \bibinfo {pages} {9537} (\bibinfo {year} {2010})},\ \Eprint
  {http://arxiv.org/abs/http://www.pnas.org/content/107/21/9537.full.pdf}
  {http://www.pnas.org/content/107/21/9537.full.pdf} \BibitemShut {NoStop}%
\bibitem [{\citenamefont {Hadzibabic}\ \emph {et~al.}(2002)\citenamefont
  {Hadzibabic}, \citenamefont {Stan}, \citenamefont {Dieckmann}, \citenamefont
  {Gupta}, \citenamefont {Zwierlein}, \citenamefont {G\"orlitz},\ and\
  \citenamefont {Ketterle}}]{Hadzibabic_2002}%
  \BibitemOpen
  \bibfield  {author} {\bibinfo {author} {\bibfnamefont {Z.}~\bibnamefont
  {Hadzibabic}}, \bibinfo {author} {\bibfnamefont {C.~A.}\ \bibnamefont
  {Stan}}, \bibinfo {author} {\bibfnamefont {K.}~\bibnamefont {Dieckmann}},
  \bibinfo {author} {\bibfnamefont {S.}~\bibnamefont {Gupta}}, \bibinfo
  {author} {\bibfnamefont {M.~W.}\ \bibnamefont {Zwierlein}}, \bibinfo {author}
  {\bibfnamefont {A.}~\bibnamefont {G\"orlitz}}, \ and\ \bibinfo {author}
  {\bibfnamefont {W.}~\bibnamefont {Ketterle}},\ }\href {\doibase
  10.1103/PhysRevLett.88.160401} {\bibfield  {journal} {\bibinfo  {journal}
  {Phys. Rev. Lett.}\ }\textbf {\bibinfo {volume} {88}},\ \bibinfo {pages}
  {160401} (\bibinfo {year} {2002})}\BibitemShut {NoStop}%
\bibitem [{\citenamefont {B\"uchler}\ and\ \citenamefont
  {Blatter}(2003)}]{Buchler_2003}%
  \BibitemOpen
  \bibfield  {author} {\bibinfo {author} {\bibfnamefont {H.~P.}\ \bibnamefont
  {B\"uchler}}\ and\ \bibinfo {author} {\bibfnamefont {G.}~\bibnamefont
  {Blatter}},\ }\href {\doibase 10.1103/PhysRevLett.91.130404} {\bibfield
  {journal} {\bibinfo  {journal} {Phys. Rev. Lett.}\ }\textbf {\bibinfo
  {volume} {91}},\ \bibinfo {pages} {130404} (\bibinfo {year}
  {2003})}\BibitemShut {NoStop}%
\bibitem [{\citenamefont {Lewenstein}\ \emph {et~al.}(2004)\citenamefont
  {Lewenstein}, \citenamefont {Santos}, \citenamefont {Baranov},\ and\
  \citenamefont {Fehrmann}}]{Lewenstein_2004}%
  \BibitemOpen
  \bibfield  {author} {\bibinfo {author} {\bibfnamefont {M.}~\bibnamefont
  {Lewenstein}}, \bibinfo {author} {\bibfnamefont {L.}~\bibnamefont {Santos}},
  \bibinfo {author} {\bibfnamefont {M.~A.}\ \bibnamefont {Baranov}}, \ and\
  \bibinfo {author} {\bibfnamefont {H.}~\bibnamefont {Fehrmann}},\ }\href
  {\doibase 10.1103/PhysRevLett.92.050401} {\bibfield  {journal} {\bibinfo
  {journal} {Phys. Rev. Lett.}\ }\textbf {\bibinfo {volume} {92}},\ \bibinfo
  {pages} {050401} (\bibinfo {year} {2004})}\BibitemShut {NoStop}%
\bibitem [{\citenamefont {B\"uchler}\ and\ \citenamefont
  {Blatter}(2004)}]{Buchler_2004}%
  \BibitemOpen
  \bibfield  {author} {\bibinfo {author} {\bibfnamefont {H.~P.}\ \bibnamefont
  {B\"uchler}}\ and\ \bibinfo {author} {\bibfnamefont {G.}~\bibnamefont
  {Blatter}},\ }\href {\doibase 10.1103/PhysRevA.69.063603} {\bibfield
  {journal} {\bibinfo  {journal} {Phys. Rev. A}\ }\textbf {\bibinfo {volume}
  {69}},\ \bibinfo {pages} {063603} (\bibinfo {year} {2004})}\BibitemShut
  {NoStop}%
\bibitem [{\citenamefont {Cramer}\ \emph {et~al.}(2004)\citenamefont {Cramer},
  \citenamefont {Eisert},\ and\ \citenamefont {Illuminati}}]{Cramer_2004}%
  \BibitemOpen
  \bibfield  {author} {\bibinfo {author} {\bibfnamefont {M.}~\bibnamefont
  {Cramer}}, \bibinfo {author} {\bibfnamefont {J.}~\bibnamefont {Eisert}}, \
  and\ \bibinfo {author} {\bibfnamefont {F.}~\bibnamefont {Illuminati}},\
  }\href {\doibase 10.1103/PhysRevLett.93.190405} {\bibfield  {journal}
  {\bibinfo  {journal} {Phys. Rev. Lett.}\ }\textbf {\bibinfo {volume} {93}},\
  \bibinfo {pages} {190405} (\bibinfo {year} {2004})}\BibitemShut {NoStop}%
\bibitem [{\citenamefont {Roth}\ and\ \citenamefont
  {Burnett}(2004)}]{Roth_2004}%
  \BibitemOpen
  \bibfield  {author} {\bibinfo {author} {\bibfnamefont {R.}~\bibnamefont
  {Roth}}\ and\ \bibinfo {author} {\bibfnamefont {K.}~\bibnamefont {Burnett}},\
  }\href {\doibase 10.1103/PhysRevA.69.021601} {\bibfield  {journal} {\bibinfo
  {journal} {Phys. Rev. A}\ }\textbf {\bibinfo {volume} {69}},\ \bibinfo
  {pages} {021601} (\bibinfo {year} {2004})}\BibitemShut {NoStop}%
\bibitem [{\citenamefont {Silber}\ \emph {et~al.}(2005)\citenamefont {Silber},
  \citenamefont {G\"unther}, \citenamefont {Marzok}, \citenamefont {Deh},
  \citenamefont {Courteille},\ and\ \citenamefont {Zimmermann}}]{Silber_2005}%
  \BibitemOpen
  \bibfield  {author} {\bibinfo {author} {\bibfnamefont {C.}~\bibnamefont
  {Silber}}, \bibinfo {author} {\bibfnamefont {S.}~\bibnamefont {G\"unther}},
  \bibinfo {author} {\bibfnamefont {C.}~\bibnamefont {Marzok}}, \bibinfo
  {author} {\bibfnamefont {B.}~\bibnamefont {Deh}}, \bibinfo {author}
  {\bibfnamefont {P.~W.}\ \bibnamefont {Courteille}}, \ and\ \bibinfo {author}
  {\bibfnamefont {C.}~\bibnamefont {Zimmermann}},\ }\href {\doibase
  10.1103/PhysRevLett.95.170408} {\bibfield  {journal} {\bibinfo  {journal}
  {Phys. Rev. Lett.}\ }\textbf {\bibinfo {volume} {95}},\ \bibinfo {pages}
  {170408} (\bibinfo {year} {2005})}\BibitemShut {NoStop}%
\bibitem [{\citenamefont {Ospelkaus}\ \emph
  {et~al.}(2006{\natexlab{a}})\citenamefont {Ospelkaus}, \citenamefont
  {Ospelkaus}, \citenamefont {Humbert}, \citenamefont {Ernst}, \citenamefont
  {Sengstock},\ and\ \citenamefont {Bongs}}]{Ospelkaus_2006a}%
  \BibitemOpen
  \bibfield  {author} {\bibinfo {author} {\bibfnamefont {C.}~\bibnamefont
  {Ospelkaus}}, \bibinfo {author} {\bibfnamefont {S.}~\bibnamefont
  {Ospelkaus}}, \bibinfo {author} {\bibfnamefont {L.}~\bibnamefont {Humbert}},
  \bibinfo {author} {\bibfnamefont {P.}~\bibnamefont {Ernst}}, \bibinfo
  {author} {\bibfnamefont {K.}~\bibnamefont {Sengstock}}, \ and\ \bibinfo
  {author} {\bibfnamefont {K.}~\bibnamefont {Bongs}},\ }\href {\doibase
  10.1103/PhysRevLett.97.120402} {\bibfield  {journal} {\bibinfo  {journal}
  {Phys. Rev. Lett.}\ }\textbf {\bibinfo {volume} {97}},\ \bibinfo {pages}
  {120402} (\bibinfo {year} {2006}{\natexlab{a}})}\BibitemShut {NoStop}%
\bibitem [{\citenamefont {Ospelkaus}\ \emph
  {et~al.}(2006{\natexlab{b}})\citenamefont {Ospelkaus}, \citenamefont
  {Ospelkaus}, \citenamefont {Humbert}, \citenamefont {Sengstock},\ and\
  \citenamefont {Bongs}}]{Ospelkaus_2006b}%
  \BibitemOpen
  \bibfield  {author} {\bibinfo {author} {\bibfnamefont {S.}~\bibnamefont
  {Ospelkaus}}, \bibinfo {author} {\bibfnamefont {C.}~\bibnamefont
  {Ospelkaus}}, \bibinfo {author} {\bibfnamefont {L.}~\bibnamefont {Humbert}},
  \bibinfo {author} {\bibfnamefont {K.}~\bibnamefont {Sengstock}}, \ and\
  \bibinfo {author} {\bibfnamefont {K.}~\bibnamefont {Bongs}},\ }\href
  {\doibase 10.1103/PhysRevLett.97.120403} {\bibfield  {journal} {\bibinfo
  {journal} {Phys. Rev. Lett.}\ }\textbf {\bibinfo {volume} {97}},\ \bibinfo
  {pages} {120403} (\bibinfo {year} {2006}{\natexlab{b}})}\BibitemShut
  {NoStop}%
\bibitem [{\citenamefont {G{\"u}nter}\ \emph {et~al.}(2006)\citenamefont
  {G{\"u}nter}, \citenamefont {St{\"o}ferle}, \citenamefont {Moritz},
  \citenamefont {K{\"o}hl},\ and\ \citenamefont {Esslinger}}]{Guenter2006}%
  \BibitemOpen
  \bibfield  {author} {\bibinfo {author} {\bibfnamefont {K.}~\bibnamefont
  {G{\"u}nter}}, \bibinfo {author} {\bibfnamefont {T.}~\bibnamefont
  {St{\"o}ferle}}, \bibinfo {author} {\bibfnamefont {H.}~\bibnamefont
  {Moritz}}, \bibinfo {author} {\bibfnamefont {M.}~\bibnamefont {K{\"o}hl}}, \
  and\ \bibinfo {author} {\bibfnamefont {T.}~\bibnamefont {Esslinger}},\ }\href
  {\doibase 10.1103/physrevlett.96.180402} {\bibfield  {journal} {\bibinfo
  {journal} {Physical Review Letters}\ }\textbf {\bibinfo {volume} {96}}
  (\bibinfo {year} {2006}),\ 10.1103/physrevlett.96.180402}\BibitemShut
  {NoStop}%
\bibitem [{\citenamefont {Powell}\ \emph {et~al.}(2005)\citenamefont {Powell},
  \citenamefont {Sachdev},\ and\ \citenamefont {B\"uchler}}]{Powell_2005}%
  \BibitemOpen
  \bibfield  {author} {\bibinfo {author} {\bibfnamefont {S.}~\bibnamefont
  {Powell}}, \bibinfo {author} {\bibfnamefont {S.}~\bibnamefont {Sachdev}}, \
  and\ \bibinfo {author} {\bibfnamefont {H.~P.}\ \bibnamefont {B\"uchler}},\
  }\href {\doibase 10.1103/PhysRevB.72.024534} {\bibfield  {journal} {\bibinfo
  {journal} {Phys. Rev. B}\ }\textbf {\bibinfo {volume} {72}},\ \bibinfo
  {pages} {024534} (\bibinfo {year} {2005})}\BibitemShut {NoStop}%
\bibitem [{\citenamefont {Ni}\ \emph {et~al.}(2008)\citenamefont {Ni},
  \citenamefont {Ospelkaus}, \citenamefont {de~Miranda}, \citenamefont {Pe'er},
  \citenamefont {Neyenhuis}, \citenamefont {Zirbel}, \citenamefont
  {Kotochigova}, \citenamefont {Julienne}, \citenamefont {Jin},\ and\
  \citenamefont {Ye}}]{Ni_2008}%
  \BibitemOpen
  \bibfield  {author} {\bibinfo {author} {\bibfnamefont {K.-K.}\ \bibnamefont
  {Ni}}, \bibinfo {author} {\bibfnamefont {S.}~\bibnamefont {Ospelkaus}},
  \bibinfo {author} {\bibfnamefont {M.~H.~G.}\ \bibnamefont {de~Miranda}},
  \bibinfo {author} {\bibfnamefont {A.}~\bibnamefont {Pe'er}}, \bibinfo
  {author} {\bibfnamefont {B.}~\bibnamefont {Neyenhuis}}, \bibinfo {author}
  {\bibfnamefont {J.~J.}\ \bibnamefont {Zirbel}}, \bibinfo {author}
  {\bibfnamefont {S.}~\bibnamefont {Kotochigova}}, \bibinfo {author}
  {\bibfnamefont {P.~S.}\ \bibnamefont {Julienne}}, \bibinfo {author}
  {\bibfnamefont {D.~S.}\ \bibnamefont {Jin}}, \ and\ \bibinfo {author}
  {\bibfnamefont {J.}~\bibnamefont {Ye}},\ }\href {\doibase
  10.1126/science.1163861} {\bibfield  {journal} {\bibinfo  {journal}
  {Science}\ }\textbf {\bibinfo {volume} {322}},\ \bibinfo {pages} {231}
  (\bibinfo {year} {2008})},\ \Eprint
  {http://arxiv.org/abs/http://www.sciencemag.org/content/322/5899/231.full.pdf}
  {http://www.sciencemag.org/content/322/5899/231.full.pdf} \BibitemShut
  {NoStop}%
\bibitem [{\citenamefont {Best}\ \emph {et~al.}(2009)\citenamefont {Best},
  \citenamefont {Will}, \citenamefont {Schneider}, \citenamefont
  {Hackerm{\"u}ller}, \citenamefont {van Oosten}, \citenamefont {Bloch},\ and\
  \citenamefont {L{\"u}hmann}}]{Best2009}%
  \BibitemOpen
  \bibfield  {author} {\bibinfo {author} {\bibfnamefont {T.}~\bibnamefont
  {Best}}, \bibinfo {author} {\bibfnamefont {S.}~\bibnamefont {Will}}, \bibinfo
  {author} {\bibfnamefont {U.}~\bibnamefont {Schneider}}, \bibinfo {author}
  {\bibfnamefont {L.}~\bibnamefont {Hackerm{\"u}ller}}, \bibinfo {author}
  {\bibfnamefont {D.}~\bibnamefont {van Oosten}}, \bibinfo {author}
  {\bibfnamefont {I.}~\bibnamefont {Bloch}}, \ and\ \bibinfo {author}
  {\bibfnamefont {D.-S.}\ \bibnamefont {L{\"u}hmann}},\ }\href {\doibase
  10.1103/physrevlett.102.030408} {\bibfield  {journal} {\bibinfo  {journal}
  {Physical Review Letters}\ }\textbf {\bibinfo {volume} {102}} (\bibinfo
  {year} {2009}),\ 10.1103/physrevlett.102.030408}\BibitemShut {NoStop}%
\bibitem [{\citenamefont {Pollet}\ \emph {et~al.}(2006)\citenamefont {Pollet},
  \citenamefont {Troyer}, \citenamefont {Van~Houcke},\ and\ \citenamefont
  {Rombouts}}]{Pollet_2006}%
  \BibitemOpen
  \bibfield  {author} {\bibinfo {author} {\bibfnamefont {L.}~\bibnamefont
  {Pollet}}, \bibinfo {author} {\bibfnamefont {M.}~\bibnamefont {Troyer}},
  \bibinfo {author} {\bibfnamefont {K.}~\bibnamefont {Van~Houcke}}, \ and\
  \bibinfo {author} {\bibfnamefont {S.~M.~A.}\ \bibnamefont {Rombouts}},\
  }\href {\doibase 10.1103/PhysRevLett.96.190402} {\bibfield  {journal}
  {\bibinfo  {journal} {Phys. Rev. Lett.}\ }\textbf {\bibinfo {volume} {96}},\
  \bibinfo {pages} {190402} (\bibinfo {year} {2006})}\BibitemShut {NoStop}%
\bibitem [{\citenamefont {Pollet}\ \emph {et~al.}(2008)\citenamefont {Pollet},
  \citenamefont {Kollath}, \citenamefont {Schollw\"ock},\ and\ \citenamefont
  {Troyer}}]{Pollet_2008}%
  \BibitemOpen
  \bibfield  {author} {\bibinfo {author} {\bibfnamefont {L.}~\bibnamefont
  {Pollet}}, \bibinfo {author} {\bibfnamefont {C.}~\bibnamefont {Kollath}},
  \bibinfo {author} {\bibfnamefont {U.}~\bibnamefont {Schollw\"ock}}, \ and\
  \bibinfo {author} {\bibfnamefont {M.}~\bibnamefont {Troyer}},\ }\href
  {\doibase 10.1103/PhysRevA.77.023608} {\bibfield  {journal} {\bibinfo
  {journal} {Phys. Rev. A}\ }\textbf {\bibinfo {volume} {77}},\ \bibinfo
  {pages} {023608} (\bibinfo {year} {2008})}\BibitemShut {NoStop}%
\bibitem [{\citenamefont {Marchetti}\ \emph {et~al.}(2008)\citenamefont
  {Marchetti}, \citenamefont {Mathy}, \citenamefont {Huse},\ and\ \citenamefont
  {Parish}}]{Parish_2008}%
  \BibitemOpen
  \bibfield  {author} {\bibinfo {author} {\bibfnamefont {F.~M.}\ \bibnamefont
  {Marchetti}}, \bibinfo {author} {\bibfnamefont {C.~J.~M.}\ \bibnamefont
  {Mathy}}, \bibinfo {author} {\bibfnamefont {D.~A.}\ \bibnamefont {Huse}}, \
  and\ \bibinfo {author} {\bibfnamefont {M.~M.}\ \bibnamefont {Parish}},\
  }\href {\doibase 10.1103/PhysRevB.78.134517} {\bibfield  {journal} {\bibinfo
  {journal} {Phys. Rev. B}\ }\textbf {\bibinfo {volume} {78}},\ \bibinfo
  {pages} {134517} (\bibinfo {year} {2008})}\BibitemShut {NoStop}%
\bibitem [{\citenamefont {Fratini}\ and\ \citenamefont
  {Pieri}(2010)}]{Fratini_2010}%
  \BibitemOpen
  \bibfield  {author} {\bibinfo {author} {\bibfnamefont {E.}~\bibnamefont
  {Fratini}}\ and\ \bibinfo {author} {\bibfnamefont {P.}~\bibnamefont
  {Pieri}},\ }\href {\doibase 10.1103/PhysRevA.81.051605} {\bibfield  {journal}
  {\bibinfo  {journal} {Phys. Rev. A}\ }\textbf {\bibinfo {volume} {81}},\
  \bibinfo {pages} {051605} (\bibinfo {year} {2010})}\BibitemShut {NoStop}%
\bibitem [{\citenamefont {Hansen}\ \emph {et~al.}(2011)\citenamefont {Hansen},
  \citenamefont {Khramov}, \citenamefont {Dowd}, \citenamefont {Jamison},
  \citenamefont {Ivanov},\ and\ \citenamefont {Gupta}}]{Hansen_2011}%
  \BibitemOpen
  \bibfield  {author} {\bibinfo {author} {\bibfnamefont {A.~H.}\ \bibnamefont
  {Hansen}}, \bibinfo {author} {\bibfnamefont {A.}~\bibnamefont {Khramov}},
  \bibinfo {author} {\bibfnamefont {W.~H.}\ \bibnamefont {Dowd}}, \bibinfo
  {author} {\bibfnamefont {A.~O.}\ \bibnamefont {Jamison}}, \bibinfo {author}
  {\bibfnamefont {V.~V.}\ \bibnamefont {Ivanov}}, \ and\ \bibinfo {author}
  {\bibfnamefont {S.}~\bibnamefont {Gupta}},\ }\href {\doibase
  10.1103/PhysRevA.84.011606} {\bibfield  {journal} {\bibinfo  {journal} {Phys.
  Rev. A}\ }\textbf {\bibinfo {volume} {84}},\ \bibinfo {pages} {011606}
  (\bibinfo {year} {2011})}\BibitemShut {NoStop}%
\bibitem [{\citenamefont {Hara}\ \emph {et~al.}(2011)\citenamefont {Hara},
  \citenamefont {Takasu}, \citenamefont {Yamaoka}, \citenamefont {Doyle},\ and\
  \citenamefont {Takahashi}}]{Hara_2011}%
  \BibitemOpen
  \bibfield  {author} {\bibinfo {author} {\bibfnamefont {H.}~\bibnamefont
  {Hara}}, \bibinfo {author} {\bibfnamefont {Y.}~\bibnamefont {Takasu}},
  \bibinfo {author} {\bibfnamefont {Y.}~\bibnamefont {Yamaoka}}, \bibinfo
  {author} {\bibfnamefont {J.~M.}\ \bibnamefont {Doyle}}, \ and\ \bibinfo
  {author} {\bibfnamefont {Y.}~\bibnamefont {Takahashi}},\ }\href {\doibase
  10.1103/PhysRevLett.106.205304} {\bibfield  {journal} {\bibinfo  {journal}
  {Phys. Rev. Lett.}\ }\textbf {\bibinfo {volume} {106}},\ \bibinfo {pages}
  {205304} (\bibinfo {year} {2011})}\BibitemShut {NoStop}%
\bibitem [{\citenamefont {Yu}\ \emph {et~al.}(2011)\citenamefont {Yu},
  \citenamefont {Zhang},\ and\ \citenamefont {Zhai}}]{Yu_2012}%
  \BibitemOpen
  \bibfield  {author} {\bibinfo {author} {\bibfnamefont {Z.-Q.}\ \bibnamefont
  {Yu}}, \bibinfo {author} {\bibfnamefont {S.}~\bibnamefont {Zhang}}, \ and\
  \bibinfo {author} {\bibfnamefont {H.}~\bibnamefont {Zhai}},\ }\href {\doibase
  10.1103/PhysRevA.83.041603} {\bibfield  {journal} {\bibinfo  {journal} {Phys.
  Rev. A}\ }\textbf {\bibinfo {volume} {83}},\ \bibinfo {pages} {041603}
  (\bibinfo {year} {2011})}\BibitemShut {NoStop}%
\bibitem [{\citenamefont {Bertaina}\ \emph {et~al.}(2013)\citenamefont
  {Bertaina}, \citenamefont {Fratini}, \citenamefont {Giorgini},\ and\
  \citenamefont {Pieri}}]{Pieri_2013}%
  \BibitemOpen
  \bibfield  {author} {\bibinfo {author} {\bibfnamefont {G.}~\bibnamefont
  {Bertaina}}, \bibinfo {author} {\bibfnamefont {E.}~\bibnamefont {Fratini}},
  \bibinfo {author} {\bibfnamefont {S.}~\bibnamefont {Giorgini}}, \ and\
  \bibinfo {author} {\bibfnamefont {P.}~\bibnamefont {Pieri}},\ }\href
  {\doibase 10.1103/PhysRevLett.110.115303} {\bibfield  {journal} {\bibinfo
  {journal} {Phys. Rev. Lett.}\ }\textbf {\bibinfo {volume} {110}},\ \bibinfo
  {pages} {115303} (\bibinfo {year} {2013})}\BibitemShut {NoStop}%
\bibitem [{\citenamefont {Guidini}\ \emph {et~al.}(2015)\citenamefont
  {Guidini}, \citenamefont {Bertaina}, \citenamefont {Galli},\ and\
  \citenamefont {Pieri}}]{Pierbiagio_2015}%
  \BibitemOpen
  \bibfield  {author} {\bibinfo {author} {\bibfnamefont {A.}~\bibnamefont
  {Guidini}}, \bibinfo {author} {\bibfnamefont {G.}~\bibnamefont {Bertaina}},
  \bibinfo {author} {\bibfnamefont {D.~E.}\ \bibnamefont {Galli}}, \ and\
  \bibinfo {author} {\bibfnamefont {P.}~\bibnamefont {Pieri}},\ }\href
  {\doibase 10.1103/PhysRevA.91.023603} {\bibfield  {journal} {\bibinfo
  {journal} {Phys. Rev. A}\ }\textbf {\bibinfo {volume} {91}},\ \bibinfo
  {pages} {023603} (\bibinfo {year} {2015})}\BibitemShut {NoStop}%
\bibitem [{\citenamefont {Ferrier-Barbut}\ \emph {et~al.}(2014)\citenamefont
  {Ferrier-Barbut}, \citenamefont {Delehaye}, \citenamefont {Laurent},
  \citenamefont {Grier}, \citenamefont {Pierce}, \citenamefont {Rem},
  \citenamefont {Chevy},\ and\ \citenamefont {Salomon}}]{Ferrier_2014}%
  \BibitemOpen
  \bibfield  {author} {\bibinfo {author} {\bibfnamefont {I.}~\bibnamefont
  {Ferrier-Barbut}}, \bibinfo {author} {\bibfnamefont {M.}~\bibnamefont
  {Delehaye}}, \bibinfo {author} {\bibfnamefont {S.}~\bibnamefont {Laurent}},
  \bibinfo {author} {\bibfnamefont {A.~T.}\ \bibnamefont {Grier}}, \bibinfo
  {author} {\bibfnamefont {M.}~\bibnamefont {Pierce}}, \bibinfo {author}
  {\bibfnamefont {B.~S.}\ \bibnamefont {Rem}}, \bibinfo {author} {\bibfnamefont
  {F.}~\bibnamefont {Chevy}}, \ and\ \bibinfo {author} {\bibfnamefont
  {C.}~\bibnamefont {Salomon}},\ }\href {\doibase 10.1126/science.1255380}
  {\bibfield  {journal} {\bibinfo  {journal} {Science}\ }\textbf {\bibinfo
  {volume} {345}},\ \bibinfo {pages} {1035} (\bibinfo {year} {2014})},\ \Eprint
  {http://arxiv.org/abs/http://www.sciencemag.org/content/345/6200/1035.full.pdf}
  {http://www.sciencemag.org/content/345/6200/1035.full.pdf} \BibitemShut
  {NoStop}%
\bibitem [{\citenamefont {Zheng}\ and\ \citenamefont
  {Zhai}(2014)}]{Zheng_2014}%
  \BibitemOpen
  \bibfield  {author} {\bibinfo {author} {\bibfnamefont {W.}~\bibnamefont
  {Zheng}}\ and\ \bibinfo {author} {\bibfnamefont {H.}~\bibnamefont {Zhai}},\
  }\href {\doibase 10.1103/PhysRevLett.113.265304} {\bibfield  {journal}
  {\bibinfo  {journal} {Phys. Rev. Lett.}\ }\textbf {\bibinfo {volume} {113}},\
  \bibinfo {pages} {265304} (\bibinfo {year} {2014})}\BibitemShut {NoStop}%
\bibitem [{\citenamefont {Kinnunen}\ and\ \citenamefont
  {Bruun}(2015)}]{Kinnunen_2015}%
  \BibitemOpen
  \bibfield  {author} {\bibinfo {author} {\bibfnamefont {J.~J.}\ \bibnamefont
  {Kinnunen}}\ and\ \bibinfo {author} {\bibfnamefont {G.~M.}\ \bibnamefont
  {Bruun}},\ }\href {\doibase 10.1103/PhysRevA.91.041605} {\bibfield  {journal}
  {\bibinfo  {journal} {Phys. Rev. A}\ }\textbf {\bibinfo {volume} {91}},\
  \bibinfo {pages} {041605} (\bibinfo {year} {2015})}\BibitemShut {NoStop}%
\bibitem [{\citenamefont {Wu}\ \emph {et~al.}(2011)\citenamefont {Wu},
  \citenamefont {Santiago}, \citenamefont {Park}, \citenamefont {Ahmadi},\ and\
  \citenamefont {Zwierlein}}]{Park_2011}%
  \BibitemOpen
  \bibfield  {author} {\bibinfo {author} {\bibfnamefont {C.-H.}\ \bibnamefont
  {Wu}}, \bibinfo {author} {\bibfnamefont {I.}~\bibnamefont {Santiago}},
  \bibinfo {author} {\bibfnamefont {J.~W.}\ \bibnamefont {Park}}, \bibinfo
  {author} {\bibfnamefont {P.}~\bibnamefont {Ahmadi}}, \ and\ \bibinfo {author}
  {\bibfnamefont {M.~W.}\ \bibnamefont {Zwierlein}},\ }\href {\doibase
  10.1103/PhysRevA.84.011601} {\bibfield  {journal} {\bibinfo  {journal} {Phys.
  Rev. A}\ }\textbf {\bibinfo {volume} {84}},\ \bibinfo {pages} {011601}
  (\bibinfo {year} {2011})}\BibitemShut {NoStop}%
\bibitem [{\citenamefont {Park}\ \emph {et~al.}(2012)\citenamefont {Park},
  \citenamefont {Wu}, \citenamefont {Santiago}, \citenamefont {Tiecke},
  \citenamefont {Will}, \citenamefont {Ahmadi},\ and\ \citenamefont
  {Zwierlein}}]{Park2012}%
  \BibitemOpen
  \bibfield  {author} {\bibinfo {author} {\bibfnamefont {J.~W.}\ \bibnamefont
  {Park}}, \bibinfo {author} {\bibfnamefont {C.-H.}\ \bibnamefont {Wu}},
  \bibinfo {author} {\bibfnamefont {I.}~\bibnamefont {Santiago}}, \bibinfo
  {author} {\bibfnamefont {T.~G.}\ \bibnamefont {Tiecke}}, \bibinfo {author}
  {\bibfnamefont {S.}~\bibnamefont {Will}}, \bibinfo {author} {\bibfnamefont
  {P.}~\bibnamefont {Ahmadi}}, \ and\ \bibinfo {author} {\bibfnamefont {M.~W.}\
  \bibnamefont {Zwierlein}},\ }\href {\doibase 10.1103/physreva.85.051602}
  {\bibfield  {journal} {\bibinfo  {journal} {Phys. Rev. A}\ }\textbf {\bibinfo
  {volume} {85}} (\bibinfo {year} {2012}),\
  10.1103/physreva.85.051602}\BibitemShut {NoStop}%
\bibitem [{\citenamefont {Park}\ \emph {et~al.}(2015)\citenamefont {Park},
  \citenamefont {Will},\ and\ \citenamefont {Zwierlein}}]{Park_2015}%
  \BibitemOpen
  \bibfield  {author} {\bibinfo {author} {\bibfnamefont {J.~W.}\ \bibnamefont
  {Park}}, \bibinfo {author} {\bibfnamefont {S.~A.}\ \bibnamefont {Will}}, \
  and\ \bibinfo {author} {\bibfnamefont {M.~W.}\ \bibnamefont {Zwierlein}},\
  }\href {\doibase 10.1103/PhysRevLett.114.205302} {\bibfield  {journal}
  {\bibinfo  {journal} {Phys. Rev. Lett.}\ }\textbf {\bibinfo {volume} {114}},\
  \bibinfo {pages} {205302} (\bibinfo {year} {2015})}\BibitemShut {NoStop}%
\bibitem [{\citenamefont {Bijlsma}\ \emph {et~al.}(2000)\citenamefont
  {Bijlsma}, \citenamefont {Heringa},\ and\ \citenamefont
  {Stoof}}]{Bijlsma2000}%
  \BibitemOpen
  \bibfield  {author} {\bibinfo {author} {\bibfnamefont {M.}~\bibnamefont
  {Bijlsma}}, \bibinfo {author} {\bibfnamefont {B.}~\bibnamefont {Heringa}}, \
  and\ \bibinfo {author} {\bibfnamefont {H.}~\bibnamefont {Stoof}},\ }\href
  {\doibase 10.1103/physreva.61.053601} {\bibfield  {journal} {\bibinfo
  {journal} {Phys. Rev. A}\ }\textbf {\bibinfo {volume} {61}} (\bibinfo {year}
  {2000}),\ 10.1103/physreva.61.053601}\BibitemShut {NoStop}%
\bibitem [{\citenamefont {Heiselberg}\ \emph {et~al.}(2000)\citenamefont
  {Heiselberg}, \citenamefont {Pethick}, \citenamefont {Smith},\ and\
  \citenamefont {Viverit}}]{Heiselberg2000}%
  \BibitemOpen
  \bibfield  {author} {\bibinfo {author} {\bibfnamefont {H.}~\bibnamefont
  {Heiselberg}}, \bibinfo {author} {\bibfnamefont {C.}~\bibnamefont {Pethick}},
  \bibinfo {author} {\bibfnamefont {H.}~\bibnamefont {Smith}}, \ and\ \bibinfo
  {author} {\bibfnamefont {L.}~\bibnamefont {Viverit}},\ }\href {\doibase
  10.1103/physrevlett.85.2418} {\bibfield  {journal} {\bibinfo  {journal}
  {Physical Review Letters}\ }\textbf {\bibinfo {volume} {85}},\ \bibinfo
  {pages} {2418–2421} (\bibinfo {year} {2000})}\BibitemShut {NoStop}%
\bibitem [{\citenamefont {Modugno}\ \emph {et~al.}(2002)\citenamefont
  {Modugno}, \citenamefont {Roati}, \citenamefont {Riboli}, \citenamefont
  {Ferlaino}, \citenamefont {Brecha},\ and\ \citenamefont
  {Inguscio}}]{Modugno_2002}%
  \BibitemOpen
  \bibfield  {author} {\bibinfo {author} {\bibfnamefont {G.}~\bibnamefont
  {Modugno}}, \bibinfo {author} {\bibfnamefont {G.}~\bibnamefont {Roati}},
  \bibinfo {author} {\bibfnamefont {F.}~\bibnamefont {Riboli}}, \bibinfo
  {author} {\bibfnamefont {F.}~\bibnamefont {Ferlaino}}, \bibinfo {author}
  {\bibfnamefont {R.~J.}\ \bibnamefont {Brecha}}, \ and\ \bibinfo {author}
  {\bibfnamefont {M.}~\bibnamefont {Inguscio}},\ }\href {\doibase
  10.1126/science.1077386} {\bibfield  {journal} {\bibinfo  {journal}
  {Science}\ }\textbf {\bibinfo {volume} {297}},\ \bibinfo {pages} {2240}
  (\bibinfo {year} {2002})},\ \Eprint
  {http://arxiv.org/abs/http://www.sciencemag.org/content/297/5590/2240.full.pdf}
  {http://www.sciencemag.org/content/297/5590/2240.full.pdf} \BibitemShut
  {NoStop}%
\bibitem [{\citenamefont {Viverit}(2002)}]{Viverit_2002a}%
  \BibitemOpen
  \bibfield  {author} {\bibinfo {author} {\bibfnamefont {L.}~\bibnamefont
  {Viverit}},\ }\href {\doibase 10.1103/PhysRevA.66.023605} {\bibfield
  {journal} {\bibinfo  {journal} {Phys. Rev. A}\ }\textbf {\bibinfo {volume}
  {66}},\ \bibinfo {pages} {023605} (\bibinfo {year} {2002})}\BibitemShut
  {NoStop}%
\bibitem [{\citenamefont {Viverit}\ and\ \citenamefont
  {Giorgini}(2002)}]{Viverit2002}%
  \BibitemOpen
  \bibfield  {author} {\bibinfo {author} {\bibfnamefont {L.}~\bibnamefont
  {Viverit}}\ and\ \bibinfo {author} {\bibfnamefont {S.}~\bibnamefont
  {Giorgini}},\ }\href {\doibase 10.1103/physreva.66.063604} {\bibfield
  {journal} {\bibinfo  {journal} {Phys. Rev. A}\ }\textbf {\bibinfo {volume}
  {66}} (\bibinfo {year} {2002}),\ 10.1103/physreva.66.063604}\BibitemShut
  {NoStop}%
\bibitem [{\citenamefont {Matera}(2003)}]{Matera_2003}%
  \BibitemOpen
  \bibfield  {author} {\bibinfo {author} {\bibfnamefont {F.}~\bibnamefont
  {Matera}},\ }\href {\doibase 10.1103/PhysRevA.68.043624} {\bibfield
  {journal} {\bibinfo  {journal} {Phys. Rev. A}\ }\textbf {\bibinfo {volume}
  {68}},\ \bibinfo {pages} {043624} (\bibinfo {year} {2003})}\BibitemShut
  {NoStop}%
\bibitem [{\citenamefont {Kagan}\ \emph {et~al.}(2004)\citenamefont {Kagan},
  \citenamefont {Brodsky}, \citenamefont {Efremov},\ and\ \citenamefont
  {Klaptsov}}]{Kagan_2004}%
  \BibitemOpen
  \bibfield  {author} {\bibinfo {author} {\bibfnamefont {M.}~\bibnamefont
  {Kagan}}, \bibinfo {author} {\bibfnamefont {I.}~\bibnamefont {Brodsky}},
  \bibinfo {author} {\bibfnamefont {D.}~\bibnamefont {Efremov}}, \ and\
  \bibinfo {author} {\bibfnamefont {A.}~\bibnamefont {Klaptsov}},\ }\href
  {\doibase 10.1134/1.1809693} {\bibfield  {journal} {\bibinfo  {journal}
  {Journal of Experimental and Theoretical Physics}\ }\textbf {\bibinfo
  {volume} {99}},\ \bibinfo {pages} {640–646} (\bibinfo {year}
  {2004})}\BibitemShut {NoStop}%
\bibitem [{\citenamefont {Illuminati}\ and\ \citenamefont
  {Albus}(2004)}]{Illuminati_2004}%
  \BibitemOpen
  \bibfield  {author} {\bibinfo {author} {\bibfnamefont {F.}~\bibnamefont
  {Illuminati}}\ and\ \bibinfo {author} {\bibfnamefont {A.}~\bibnamefont
  {Albus}},\ }\href {\doibase 10.1103/PhysRevLett.93.090406} {\bibfield
  {journal} {\bibinfo  {journal} {Phys. Rev. Lett.}\ }\textbf {\bibinfo
  {volume} {93}},\ \bibinfo {pages} {090406} (\bibinfo {year}
  {2004})}\BibitemShut {NoStop}%
\bibitem [{\citenamefont {Wang}(2006)}]{Wang_2006}%
  \BibitemOpen
  \bibfield  {author} {\bibinfo {author} {\bibfnamefont {D.-W.}\ \bibnamefont
  {Wang}},\ }\href {\doibase 10.1103/PhysRevLett.96.140404} {\bibfield
  {journal} {\bibinfo  {journal} {Phys. Rev. Lett.}\ }\textbf {\bibinfo
  {volume} {96}},\ \bibinfo {pages} {140404} (\bibinfo {year}
  {2006})}\BibitemShut {NoStop}%
\bibitem [{\citenamefont {Kalas}\ \emph {et~al.}(2008)\citenamefont {Kalas},
  \citenamefont {Balatsky},\ and\ \citenamefont {Mozyrsky}}]{Kalas_2008}%
  \BibitemOpen
  \bibfield  {author} {\bibinfo {author} {\bibfnamefont {R.~M.}\ \bibnamefont
  {Kalas}}, \bibinfo {author} {\bibfnamefont {A.~V.}\ \bibnamefont {Balatsky}},
  \ and\ \bibinfo {author} {\bibfnamefont {D.}~\bibnamefont {Mozyrsky}},\
  }\href {\doibase 10.1103/PhysRevB.78.184513} {\bibfield  {journal} {\bibinfo
  {journal} {Phys. Rev. B}\ }\textbf {\bibinfo {volume} {78}},\ \bibinfo
  {pages} {184513} (\bibinfo {year} {2008})}\BibitemShut {NoStop}%
\bibitem [{\citenamefont {Enss}\ and\ \citenamefont
  {Zwerger}(2009)}]{Zwerger_2009}%
  \BibitemOpen
  \bibfield  {author} {\bibinfo {author} {\bibfnamefont {T.}~\bibnamefont
  {Enss}}\ and\ \bibinfo {author} {\bibfnamefont {W.}~\bibnamefont {Zwerger}},\
  }\href {\doibase 10.1140/epjb/e2009-00005-y} {\bibfield  {journal} {\bibinfo
  {journal} {The European Physical Journal B}\ }\textbf {\bibinfo {volume}
  {68}},\ \bibinfo {pages} {383–389} (\bibinfo {year} {2009})}\BibitemShut
  {NoStop}%
\bibitem [{\citenamefont {Anders}\ \emph {et~al.}(2012)\citenamefont {Anders},
  \citenamefont {Werner}, \citenamefont {Troyer}, \citenamefont {Sigrist},\
  and\ \citenamefont {Pollet}}]{Anders2012}%
  \BibitemOpen
  \bibfield  {author} {\bibinfo {author} {\bibfnamefont {P.}~\bibnamefont
  {Anders}}, \bibinfo {author} {\bibfnamefont {P.}~\bibnamefont {Werner}},
  \bibinfo {author} {\bibfnamefont {M.}~\bibnamefont {Troyer}}, \bibinfo
  {author} {\bibfnamefont {M.}~\bibnamefont {Sigrist}}, \ and\ \bibinfo
  {author} {\bibfnamefont {L.}~\bibnamefont {Pollet}},\ }\href {\doibase
  10.1103/physrevlett.109.206401} {\bibfield  {journal} {\bibinfo  {journal}
  {Physical Review Letters}\ }\textbf {\bibinfo {volume} {109}} (\bibinfo
  {year} {2012}),\ 10.1103/physrevlett.109.206401}\BibitemShut {NoStop}%
\bibitem [{\citenamefont {Bukov}\ and\ \citenamefont
  {Pollet}(2014)}]{Bukov2014}%
  \BibitemOpen
  \bibfield  {author} {\bibinfo {author} {\bibfnamefont {M.}~\bibnamefont
  {Bukov}}\ and\ \bibinfo {author} {\bibfnamefont {L.}~\bibnamefont {Pollet}},\
  }\href {\doibase 10.1103/physrevb.89.094502} {\bibfield  {journal} {\bibinfo
  {journal} {Phys. Rev. B}\ }\textbf {\bibinfo {volume} {89}} (\bibinfo {year}
  {2014}),\ 10.1103/physrevb.89.094502}\BibitemShut {NoStop}%
\bibitem [{\citenamefont {Mathey}\ \emph {et~al.}(2006)\citenamefont {Mathey},
  \citenamefont {Tsai},\ and\ \citenamefont {Neto}}]{Mathey2006}%
  \BibitemOpen
  \bibfield  {author} {\bibinfo {author} {\bibfnamefont {L.}~\bibnamefont
  {Mathey}}, \bibinfo {author} {\bibfnamefont {S.-W.}\ \bibnamefont {Tsai}}, \
  and\ \bibinfo {author} {\bibfnamefont {A.}~\bibnamefont {Neto}},\ }\href
  {\doibase 10.1103/physrevlett.97.030601} {\bibfield  {journal} {\bibinfo
  {journal} {Physical Review Letters}\ }\textbf {\bibinfo {volume} {97}}
  (\bibinfo {year} {2006}),\ 10.1103/physrevlett.97.030601}\BibitemShut
  {NoStop}%
\bibitem [{\citenamefont {Hettler}\ \emph {et~al.}(1998)\citenamefont
  {Hettler}, \citenamefont {Tahvildar-Zadeh}, \citenamefont {Jarrell},
  \citenamefont {Pruschke},\ and\ \citenamefont {Krishnamurthy}}]{Hettler1998}%
  \BibitemOpen
  \bibfield  {author} {\bibinfo {author} {\bibfnamefont {M.}~\bibnamefont
  {Hettler}}, \bibinfo {author} {\bibfnamefont {A.}~\bibnamefont
  {Tahvildar-Zadeh}}, \bibinfo {author} {\bibfnamefont {M.}~\bibnamefont
  {Jarrell}}, \bibinfo {author} {\bibfnamefont {T.}~\bibnamefont {Pruschke}}, \
  and\ \bibinfo {author} {\bibfnamefont {H.}~\bibnamefont {Krishnamurthy}},\
  }\href {\doibase 10.1103/physrevb.58.r7475} {\bibfield  {journal} {\bibinfo
  {journal} {Phys. Rev. B}\ }\textbf {\bibinfo {volume} {58}},\ \bibinfo
  {pages} {R7475–R7479} (\bibinfo {year} {1998})}\BibitemShut {NoStop}%
\bibitem [{\citenamefont {Jarrell}\ \emph {et~al.}(2001)\citenamefont
  {Jarrell}, \citenamefont {Maier}, \citenamefont {Huscroft},\ and\
  \citenamefont {Moukouri}}]{Jarrell_2001a}%
  \BibitemOpen
  \bibfield  {author} {\bibinfo {author} {\bibfnamefont {M.}~\bibnamefont
  {Jarrell}}, \bibinfo {author} {\bibfnamefont {T.}~\bibnamefont {Maier}},
  \bibinfo {author} {\bibfnamefont {C.}~\bibnamefont {Huscroft}}, \ and\
  \bibinfo {author} {\bibfnamefont {S.}~\bibnamefont {Moukouri}},\ }\href
  {\doibase 10.1103/physrevb.64.195130} {\bibfield  {journal} {\bibinfo
  {journal} {Phys. Rev. B}\ }\textbf {\bibinfo {volume} {64}} (\bibinfo {year}
  {2001}),\ 10.1103/physrevb.64.195130}\BibitemShut {NoStop}%
\bibitem [{\citenamefont {Maier}\ \emph
  {et~al.}(2005{\natexlab{a}})\citenamefont {Maier}, \citenamefont {Jarrell},
  \citenamefont {Pruschke},\ and\ \citenamefont {Hettler}}]{Maier2005}%
  \BibitemOpen
  \bibfield  {author} {\bibinfo {author} {\bibfnamefont {T.}~\bibnamefont
  {Maier}}, \bibinfo {author} {\bibfnamefont {M.}~\bibnamefont {Jarrell}},
  \bibinfo {author} {\bibfnamefont {T.}~\bibnamefont {Pruschke}}, \ and\
  \bibinfo {author} {\bibfnamefont {M.}~\bibnamefont {Hettler}},\ }\href
  {\doibase 10.1103/revmodphys.77.1027} {\bibfield  {journal} {\bibinfo
  {journal} {Rev. Mod. Phys.}\ }\textbf {\bibinfo {volume} {77}},\ \bibinfo
  {pages} {1027–1080} (\bibinfo {year} {2005}{\natexlab{a}})}\BibitemShut
  {NoStop}%
\bibitem [{\citenamefont {Rubtsov}\ and\ \citenamefont
  {Lichtenstein}(2004)}]{Rubtsov2004}%
  \BibitemOpen
  \bibfield  {author} {\bibinfo {author} {\bibfnamefont {A.~N.}\ \bibnamefont
  {Rubtsov}}\ and\ \bibinfo {author} {\bibfnamefont {A.~I.}\ \bibnamefont
  {Lichtenstein}},\ }\href {\doibase 10.1134/1.1800216} {\bibfield  {journal}
  {\bibinfo  {journal} {Journal of Experimental and Theoretical Physics
  Letters}\ }\textbf {\bibinfo {volume} {80}},\ \bibinfo {pages} {61–65}
  (\bibinfo {year} {2004})}\BibitemShut {NoStop}%
\bibitem [{\citenamefont {Rubtsov}\ \emph {et~al.}(2005)\citenamefont
  {Rubtsov}, \citenamefont {Savkin},\ and\ \citenamefont
  {Lichtenstein}}]{Rubtsov2005}%
  \BibitemOpen
  \bibfield  {author} {\bibinfo {author} {\bibfnamefont {A.}~\bibnamefont
  {Rubtsov}}, \bibinfo {author} {\bibfnamefont {V.}~\bibnamefont {Savkin}}, \
  and\ \bibinfo {author} {\bibfnamefont {A.}~\bibnamefont {Lichtenstein}},\
  }\href {\doibase 10.1103/physrevb.72.035122} {\bibfield  {journal} {\bibinfo
  {journal} {Phys. Rev. B}\ }\textbf {\bibinfo {volume} {72}} (\bibinfo {year}
  {2005}),\ 10.1103/physrevb.72.035122}\BibitemShut {NoStop}%
\bibitem [{\citenamefont {Assaad}\ and\ \citenamefont
  {Lang}(2007)}]{Assaad2007}%
  \BibitemOpen
  \bibfield  {author} {\bibinfo {author} {\bibfnamefont {F.}~\bibnamefont
  {Assaad}}\ and\ \bibinfo {author} {\bibfnamefont {T.}~\bibnamefont {Lang}},\
  }\href {\doibase 10.1103/physrevb.76.035116} {\bibfield  {journal} {\bibinfo
  {journal} {Phys. Rev. B}\ }\textbf {\bibinfo {volume} {76}} (\bibinfo {year}
  {2007}),\ 10.1103/physrevb.76.035116}\BibitemShut {NoStop}%
\bibitem [{\citenamefont {Maier}\ \emph
  {et~al.}(2005{\natexlab{b}})\citenamefont {Maier}, \citenamefont {Jarrell},
  \citenamefont {Schulthess}, \citenamefont {Kent},\ and\ \citenamefont
  {White}}]{Maier2005a}%
  \BibitemOpen
  \bibfield  {author} {\bibinfo {author} {\bibfnamefont {T.}~\bibnamefont
  {Maier}}, \bibinfo {author} {\bibfnamefont {M.}~\bibnamefont {Jarrell}},
  \bibinfo {author} {\bibfnamefont {T.}~\bibnamefont {Schulthess}}, \bibinfo
  {author} {\bibfnamefont {P.}~\bibnamefont {Kent}}, \ and\ \bibinfo {author}
  {\bibfnamefont {J.}~\bibnamefont {White}},\ }\href {\doibase
  10.1103/physrevlett.95.237001} {\bibfield  {journal} {\bibinfo  {journal}
  {Physical Review Letters}\ }\textbf {\bibinfo {volume} {95}} (\bibinfo {year}
  {2005}{\natexlab{b}}),\ 10.1103/physrevlett.95.237001}\BibitemShut {NoStop}%
\bibitem [{\citenamefont {Abrikosov}(1975)}]{Abrikosov1975}%
  \BibitemOpen
  \bibfield  {author} {\bibinfo {author} {\bibfnamefont {A.~A.}\ \bibnamefont
  {Abrikosov}},\ }\href@noop {} {\emph {\bibinfo {title} {Methods of Quantum
  Field Theory in Statistical Physics (Dover Books on Physics)}}},\ \bibinfo
  {edition} {revised}\ ed.\ (\bibinfo  {publisher} {Dover Publications},\
  \bibinfo {year} {1975})\BibitemShut {NoStop}%
\bibitem [{\citenamefont {Migdal}(1958)}]{Mig_58}%
  \BibitemOpen
  \bibfield  {author} {\bibinfo {author} {\bibfnamefont {A.~B.}\ \bibnamefont
  {Migdal}},\ }\href@noop {} {\bibfield  {journal} {\bibinfo  {journal} {Sov.
  Phys. JETP}\ }\textbf {\bibinfo {volume} {7}},\ \bibinfo {pages} {996}
  (\bibinfo {year} {1958})}\BibitemShut {NoStop}%
\bibitem [{\citenamefont {Eliashberg}(1960)}]{Eli_60}%
  \BibitemOpen
  \bibfield  {author} {\bibinfo {author} {\bibfnamefont {G.~M.}\ \bibnamefont
  {Eliashberg}},\ }\href@noop {} {\bibfield  {journal} {\bibinfo  {journal}
  {Sov. Phys. JETP}\ }\textbf {\bibinfo {volume} {11}} (\bibinfo {year}
  {1960})}\BibitemShut {NoStop}%
\bibitem [{\citenamefont {Carbotte}(1990)}]{Carbotte_1990}%
  \BibitemOpen
  \bibfield  {author} {\bibinfo {author} {\bibfnamefont {J.~P.}\ \bibnamefont
  {Carbotte}},\ }\href {\doibase 10.1103/RevModPhys.62.1027} {\bibfield
  {journal} {\bibinfo  {journal} {Rev. Mod. Phys.}\ }\textbf {\bibinfo {volume}
  {62}},\ \bibinfo {pages} {1027} (\bibinfo {year} {1990})}\BibitemShut
  {NoStop}%
\bibitem [{\citenamefont {Marsiglio}\ and\ \citenamefont
  {Carbotte}(2008)}]{Springer_Superconductivity_ep}%
  \BibitemOpen
  \bibfield  {author} {\bibinfo {author} {\bibfnamefont {F.}~\bibnamefont
  {Marsiglio}}\ and\ \bibinfo {author} {\bibfnamefont {J.}~\bibnamefont
  {Carbotte}},\ }in\ \href {\doibase 10.1007/978-3-540-73253-2_3}
  {{\emph {\bibinfo {booktitle}
  {{Superconductivity}}}}},\ \bibinfo {editor} {edited by\ \bibinfo {editor}
  {\bibfnamefont {K.}~\bibnamefont {Bennemann}}\ and\ \bibinfo {editor}
  {\bibfnamefont {J.}~\bibnamefont {Ketterson}}}\ (\bibinfo  {publisher}
  {Springer Berlin Heidelberg},\ \bibinfo {year} {2008})\ p.\ \bibinfo {pages}
  {73–162}\BibitemShut {NoStop}%
\bibitem [{\citenamefont {Bauer}\ \emph
  {et~al.}(2011{\natexlab{a}})\citenamefont {Bauer}, \citenamefont {Han},\ and\
  \citenamefont {Gunnarsson}}]{Bauer_2011}%
  \BibitemOpen
  \bibfield  {author} {\bibinfo {author} {\bibfnamefont {J.}~\bibnamefont
  {Bauer}}, \bibinfo {author} {\bibfnamefont {J.~E.}\ \bibnamefont {Han}}, \
  and\ \bibinfo {author} {\bibfnamefont {O.}~\bibnamefont {Gunnarsson}},\
  }\href {\doibase 10.1103/PhysRevB.84.184531} {\bibfield  {journal} {\bibinfo
  {journal} {Phys. Rev. B}\ }\textbf {\bibinfo {volume} {84}},\ \bibinfo
  {pages} {184531} (\bibinfo {year} {2011}{\natexlab{a}})}\BibitemShut
  {NoStop}%
\bibitem [{\citenamefont {Margine}\ and\ \citenamefont
  {Giustino}(2013)}]{Margine_2013}%
  \BibitemOpen
  \bibfield  {author} {\bibinfo {author} {\bibfnamefont {E.~R.}\ \bibnamefont
  {Margine}}\ and\ \bibinfo {author} {\bibfnamefont {F.}~\bibnamefont
  {Giustino}},\ }\href {\doibase 10.1103/PhysRevB.87.024505} {\bibfield
  {journal} {\bibinfo  {journal} {Phys. Rev. B}\ }\textbf {\bibinfo {volume}
  {87}},\ \bibinfo {pages} {024505} (\bibinfo {year} {2013})}\BibitemShut
  {NoStop}%
\bibitem [{\citenamefont {Bauer}\ \emph
  {et~al.}(2011{\natexlab{b}})\citenamefont {Bauer}, \citenamefont {Carr},
  \citenamefont {Evertz}, \citenamefont {Feiguin}, \citenamefont {Freire},
  \citenamefont {Fuchs}, \citenamefont {Gamper}, \citenamefont {Gukelberger},
  \citenamefont {Gull}, \citenamefont {Guertler}, \citenamefont {Hehn},
  \citenamefont {Igarashi}, \citenamefont {Isakov}, \citenamefont {Koop},
  \citenamefont {Ma}, \citenamefont {Mates}, \citenamefont {Matsuo},
  \citenamefont {Parcollet}, \citenamefont {Paw{\l}owski}, \citenamefont
  {Picon}, \citenamefont {Pollet}, \citenamefont {Santos}, \citenamefont
  {Scarola}, \citenamefont {Schollw{\"o}ck}, \citenamefont {Silva},
  \citenamefont {Surer}, \citenamefont {Todo}, \citenamefont {Trebst},
  \citenamefont {Troyer}, \citenamefont {Wall}, \citenamefont {Werner},\ and\
  \citenamefont {Wessel}}]{ALPS}%
  \BibitemOpen
  \bibfield  {author} {\bibinfo {author} {\bibfnamefont {B.}~\bibnamefont
  {Bauer}}, \bibinfo {author} {\bibfnamefont {L.~D.}\ \bibnamefont {Carr}},
  \bibinfo {author} {\bibfnamefont {H.~G.}\ \bibnamefont {Evertz}}, \bibinfo
  {author} {\bibfnamefont {A.}~\bibnamefont {Feiguin}}, \bibinfo {author}
  {\bibfnamefont {J.}~\bibnamefont {Freire}}, \bibinfo {author} {\bibfnamefont
  {S.}~\bibnamefont {Fuchs}}, \bibinfo {author} {\bibfnamefont
  {L.}~\bibnamefont {Gamper}}, \bibinfo {author} {\bibfnamefont
  {J.}~\bibnamefont {Gukelberger}}, \bibinfo {author} {\bibfnamefont
  {E.}~\bibnamefont {Gull}}, \bibinfo {author} {\bibfnamefont {S.}~\bibnamefont
  {Guertler}}, \bibinfo {author} {\bibfnamefont {A.}~\bibnamefont {Hehn}},
  \bibinfo {author} {\bibfnamefont {R.}~\bibnamefont {Igarashi}}, \bibinfo
  {author} {\bibfnamefont {S.~V.}\ \bibnamefont {Isakov}}, \bibinfo {author}
  {\bibfnamefont {D.}~\bibnamefont {Koop}}, \bibinfo {author} {\bibfnamefont
  {P.~N.}\ \bibnamefont {Ma}}, \bibinfo {author} {\bibfnamefont
  {P.}~\bibnamefont {Mates}}, \bibinfo {author} {\bibfnamefont
  {H.}~\bibnamefont {Matsuo}}, \bibinfo {author} {\bibfnamefont
  {O.}~\bibnamefont {Parcollet}}, \bibinfo {author} {\bibfnamefont
  {G.}~\bibnamefont {Paw{\l}owski}}, \bibinfo {author} {\bibfnamefont {J.~D.}\
  \bibnamefont {Picon}}, \bibinfo {author} {\bibfnamefont {L.}~\bibnamefont
  {Pollet}}, \bibinfo {author} {\bibfnamefont {E.}~\bibnamefont {Santos}},
  \bibinfo {author} {\bibfnamefont {V.~W.}\ \bibnamefont {Scarola}}, \bibinfo
  {author} {\bibfnamefont {U.}~\bibnamefont {Schollw{\"o}ck}}, \bibinfo
  {author} {\bibfnamefont {C.}~\bibnamefont {Silva}}, \bibinfo {author}
  {\bibfnamefont {B.}~\bibnamefont {Surer}}, \bibinfo {author} {\bibfnamefont
  {S.}~\bibnamefont {Todo}}, \bibinfo {author} {\bibfnamefont {S.}~\bibnamefont
  {Trebst}}, \bibinfo {author} {\bibfnamefont {M.}~\bibnamefont {Troyer}},
  \bibinfo {author} {\bibfnamefont {M.~L.}\ \bibnamefont {Wall}}, \bibinfo
  {author} {\bibfnamefont {P.}~\bibnamefont {Werner}}, \ and\ \bibinfo {author}
  {\bibfnamefont {S.}~\bibnamefont {Wessel}},\ }\href
  {http://stacks.iop.org/1742-5468/2011/i=05/a=P05001} {\bibfield  {journal}
  {\bibinfo  {journal} {Journal of Statistical Mechanics: Theory and
  Experiment}\ }\textbf {\bibinfo {volume} {2011}},\ \bibinfo {pages} {P05001}
  (\bibinfo {year} {2011}{\natexlab{b}})}\BibitemShut {NoStop}%
\bibitem [{\citenamefont {{Gull}}\ \emph {et~al.}(2011)\citenamefont {{Gull}},
  \citenamefont {{Werner}}, \citenamefont {{Fuchs}}, \citenamefont {{Surer}},
  \citenamefont {{Pruschke}},\ and\ \citenamefont {{Troyer}}}]{ALPS_DMFT}%
  \BibitemOpen
  \bibfield  {author} {\bibinfo {author} {\bibfnamefont {E.}~\bibnamefont
  {{Gull}}}, \bibinfo {author} {\bibfnamefont {P.}~\bibnamefont {{Werner}}},
  \bibinfo {author} {\bibfnamefont {S.}~\bibnamefont {{Fuchs}}}, \bibinfo
  {author} {\bibfnamefont {B.}~\bibnamefont {{Surer}}}, \bibinfo {author}
  {\bibfnamefont {T.}~\bibnamefont {{Pruschke}}}, \ and\ \bibinfo {author}
  {\bibfnamefont {M.}~\bibnamefont {{Troyer}}},\ }\href {\doibase
  10.1016/j.cpc.2010.12.050} {\bibfield  {journal} {\bibinfo  {journal}
  {Computer Physics Communications}\ }\textbf {\bibinfo {volume} {182}},\
  \bibinfo {pages} {1078} (\bibinfo {year} {2011})}\BibitemShut {NoStop}%
\end{thebibliography}%

\end{document}